# Sonic metamaterials: reflection on the role of topology on dispersion surface morphology

V. N. Gorshkov [a*], N. Navadeh [b], P. Sareh [b], V. V. Tereshchuk [a], A. S. Fallah [b]

[a] Building 7, Department of Physics, National Technical University of Ukraine- Igor Sikorsky Kiev Polytechnic Institute, 37 Peremogy Avenue, Kiev-56, 03056, Ukraine

[b] ACEX Building, Department of Aeronautics, South Kensington Campus, Imperial College London, London SW7 2AZ, UK

**Abstract**

Investigating dispersion surface morphology of sonic metamaterials is crucial in providing information on related phenomena as inertial coupling, acoustic transparency, polarisation, and absorption. In the present study, we look into frequency surface morphology of two-dimensional (2D) metamaterials of $K_{3,3}$ and $K_6$ topologies. The elastic structures under consideration consist of the same substratum lattice points and form a pair of sublattices with hexagonal symmetry. We show that, through introducing universal localised mass-in-mass phononic microstructures at lattice points, six single optical frequency-surfaces can be formed with required properties including negative group velocity. Splitting the frequency-surfaces is based on the classical analog of the quantum phenomenon of 'energy-level repulsion', which can be achieved only through internal anisotropy of the nodes and allows us to obtain different frequency band gaps.

1. Introduction

Sonic metamaterials exhibit response characteristics not observed in natural materials, and signify certain elastodynamic features when considering vibration mitigation, wave manipulation, or sound attenuation. One important attribute of sonic metamaterials is their susceptibility to being tailored as acoustic filters. Associated properties such as negative effective mass density, negative effective elastic modulus, and the possibility of gap formation at certain frequencies allow sonic metamaterials to be used in acoustic imaging, sound wave control and vibration shielding, sonic cloaking, and impact and blast-wave mitigation.

Sonic metamaterials and lattices have been the subject of some contemporary research in order to investigate different aspects of these characteristics [1-12]. First sonic metamaterials appeared rather recently (in the year 2000). Since then, scientists have conducted a lot of research in order to explore the potentially beneficial properties of this class of metamaterials. Liu et al [13] fabricated sonic crystals with negative elastic constants which acted as a wave reflector in the vicinity of resonance frequencies. In another study, Yao et al [14] experimentally studied the 1D spring-mass model originally proposed by Milton and Willis [15]. As a result, the negative effective mass was attained which represented the situation when the internal sphere was not moving in phase with the outer bar. The effect of negative effective density can be presented by a lattice which consists of mass-in-mass units. Using this model, Huang and Sun [16] presented a metamaterial with negative effective mass density which created a band gap at a frequency close to local resonance. It was also shown that, by varying the internal parameters, one can easily shift the range of band gap frequencies which made this metamaterial efficient in blocking vibrations and low-frequency sound. Li and Chan [17] studied doubly negative acoustic metamaterials in which concurrent negative effective density and bulk modulus were obtained. Their double-negative acoustic system is an acoustic analogue of Veselago's medium in electromagnetism [18-20], and shares with it many principle features, as negative refractive index, as a consequence of its microstructural composition. In these applications metamaterials were used as filters and noise cancelling devices.

* To whom correspondence should be addressed:
Tel.: +38 (068) 322-8293
Email: vn.gorshkov@gmail.com (V.N. Gorshkov)

In a different vein, sonic metamaterials may be used as protective systems. It is evident that, for instance, blast waves are detrimental to critical systems and perilous for humans, causing severe damage to internal organs, especially to lungs (pulmonary contusion) and the hollow organs of the gastronomical tract. That is why it is crucial to design materials capable of strongly attenuating blast wave propagation. In a recent study, Tan et al [21] looked into the effect of negative effective mass on blast-wave impact mitigation. In order to demonstrate blast-wave mitigation, they used a 1D system with single and double resonators and subjected it to a blast pulse. The results of their numerical analyses showed a large amount of reflection from the metamaterial. The waves which eventually passed through the metamaterial possessed much smaller amplitudes compared to the waves which pass through conventional structures. Dynamic load mitigation using negative effective mass structures was also discussed by Manimala et al [22]. As a result, an isolator with the continuous bandwidth of isolation over a frequency range of approximately 4.5 Hz and a 98% isolation near the local resonance frequency was presented. The authors emphasised that adopting analogous negative effective density structures in the design of infrastructure building-blocks made them resilient to broadband impact-type loadings. Multi-band wave filtering in bio-inspired composites was discussed by Chen and Wang [23]. They investigated elastic wave propagation in two types of material architecture. They found out that low frequency band gaps were attributed to Bragg scattering while high frequency pass and stop bands were ascribed due to the manifestation of trapped and transmitted waveguides. The latter forms the foundation of the present study while the former falls beyond its scope.

Although theoretical analyses imply that the unique properties of sonic metamaterials make their implementation extremely effective in solving various problems connected with wave propagation, their fabrication remained a challenge for a long time. Difficulties were associated with the creation of resonators which would have a wide range of sizes, masses, and more than one constituent. With the advent of 3D printing technologies the methods of microfabrication have gone through a significant breakthrough in construction of sonic metamaterials. In a recent study, Qureshi et al [24] used the Objet Eden260v 3D printer and the photo-polymerisation technique to fabricate the cantilever-in-mass structure with a single material constituent. Using a dual-extruding 3D printing device Gandy and Niu [25] have already manufactured a metamaterial acoustic lens. Also, using the same technology, researchers have recently managed to manufacture metastructures for low-frequency and broadband vibration absorption [26].

It is a well-known fact that the constitutive mechanical behaviour of a sonic metamaterial is not determined by its atomic structure but rather by its unit cell. The unit cell is, in its own right, determined by lattice constants, microstructural composition, as well as the topology of the system. For a given topology, generalised anisotropic micro-continuum models could be proposed to represent the equivalent homogeneous system with effective parameters (see e.g. [27]). Topological insulators have been studied by researchers with the objective of obtaining their band gap properties [11]. Furthermore, depending on the length scale and objectives different mechanisms of gap formation can be observed and different systems proposed. Chen and Wang [28] studied overlapping locally resonant and Bragg band gaps in acoustic metamaterials exhibiting simultaneous wave filtering capability and enhanced mechanical properties. They studied topological influence on the coupling between the two phenomena and suggested the design could simultaneously absorb acoustic and guide elastic waves.

We investigate frequency surface morphology and band gap structure for 2D sonic metamaterials of $K_{3,3}$ and $K_6$ topologies. Both metamaterials are presented with the honeycomb lattice constructed of locally resonant mas-in mass units (nodes) connected by springs in the corresponding topology. The unit cell of the lattice contains two nodes (Fig. 1a), which may, in general, have different physical characteristics. Thus, the system under consideration could be considered as a pair of sublattices with hexagonal symmetry consisting of either red (A) or blue (B) nodes (with masses of $M_a$ and $M_b$, respectively).

When the topology is that of $K_{3,3}$, each A/B node interacts only with six B/A-nodes placed within a circle of radius $2a$ ($a$ being the shortest distance between nodes) and centered at that node– See Fig. 1b and Fig. 2b. For the case of $K_6$ topology the additional interactions between nodes of the same type are taken into account - See Fig. 1c and Fig. 2a. (We shall assume that the stiffness of a massless spring is inversely proportional to its length). In both cases, the internal nodal structures are introduced such that they represent three configurations, each with its own orientation and number of internal springs – Fig. 3.

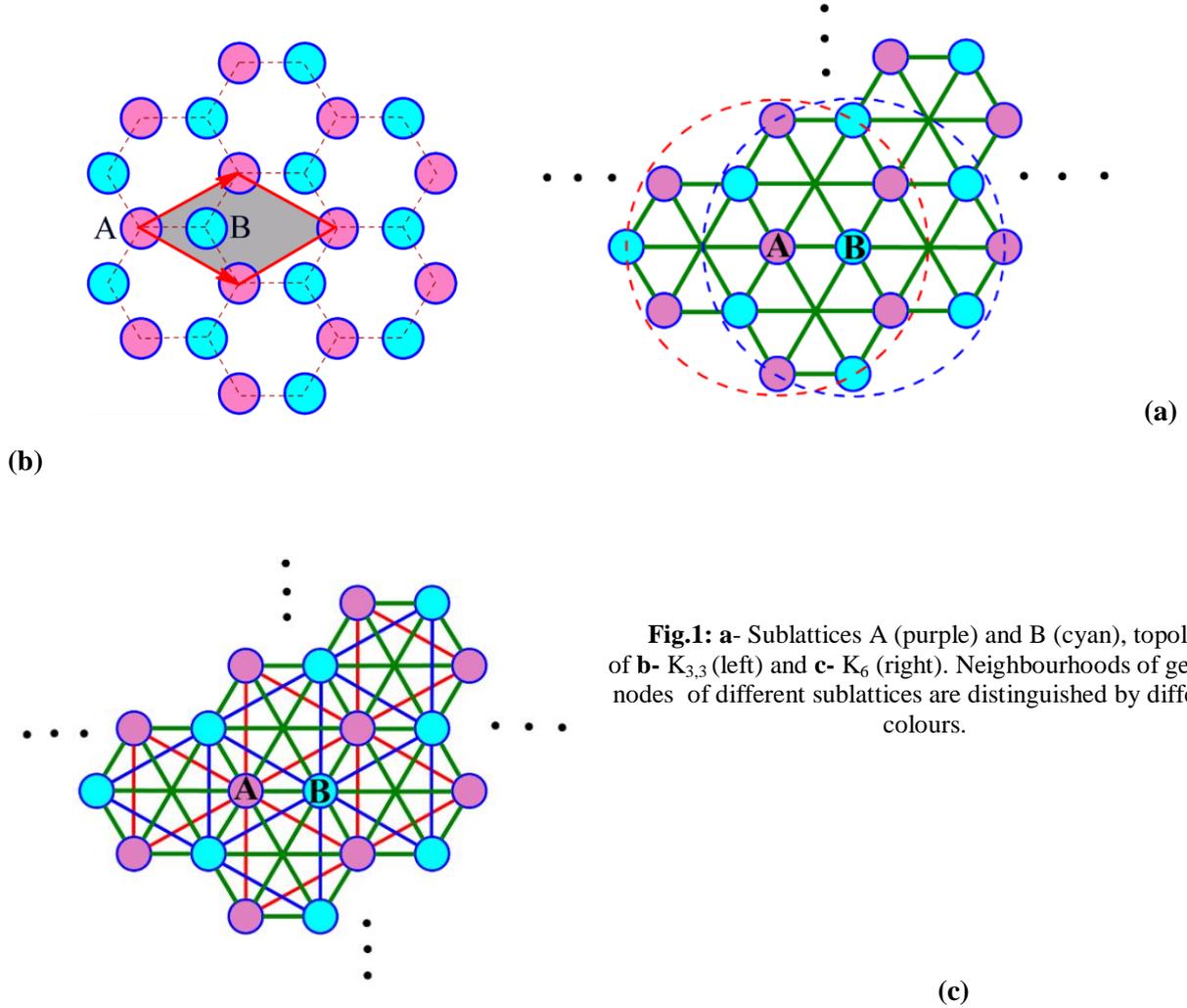

**Fig.1:** **a**- Sublattices A (purple) and B (cyan), topologies of **b-** $K_{3,3}$ (left) and **c-** $K_6$ (right). Neighbourhoods of generic nodes of different sublattices are distinguished by different colours.

The mechanisms of band gap formation, which are analyzed in detail in the sequel, can be briefly presented as follows: In the general case, the solution of the problem is the set of eight frequency-surfaces $\omega_n^2(\boldsymbol{k})$ comprising two low frequency acoustic modes ($n = 1,2$, $\omega_n^2(\boldsymbol{k} = \boldsymbol{0}) = 0$) and six optical modes ($n = 3,4,\ldots,8$, $\omega_n^2(\boldsymbol{k} = \boldsymbol{0}) \neq 0$). To form the six band gaps over the first BZ, the optical frequencies $\omega_n^2(\boldsymbol{k} = \boldsymbol{0})$ must be noticeably separated from each other at least at its center. This can be realized by using anisotropic acoustic properties of the internal structures of nodes, which may be regarded as isolated oscillators imbedded in the self-consistent elastic-force field produced by external springs. This field has three possibilities to be formed, each realization of which can be boiled down to "free" vibrations of the single node with some specific effective shell mass, $\widehat{M}(\boldsymbol{k})$. As a result, at $k = 0$ *each of the initial anisotropic eigen-frequencies,* $\lambda_x^2$, $\lambda_y^2$, of the single node (supposed to be the same for the A- and B-sublattices) transforms into three modes $\omega_l^2(\boldsymbol{k})$ that correspond to these different effective masses of the node shell.

Summarizing the above, at $k = 0$ the set of the effective masses may be presented as $\{M, \widehat{M}_x^{(+)}, \widehat{M}_y^{(+)} < M, \widehat{M}_x^{(-)}, \widehat{M}_y^{(-)} < 0\}$, where $M$ is the real nodal mass, and the set of frequencies as:

$$S_\omega = \{\lambda_x^2, \lambda_y^2, \ \omega^2(\widehat{M}_x^{(+)}) > \lambda_x^2, \ \omega^2\left(\widehat{M}_y^{(+)}\right) > \lambda_y^2, \ \omega^2(\widehat{M}_x^{(-)}) < \lambda_x^2, \ \omega^2(\widehat{M}_y^{(-)}) < \lambda_y^2\}.$$

The effects of negative masses, frequency-surface separation and band gaps formation intensify if the internal masses, $m_{a,b}$, are of the order of or greater than the sell masses. In the case of the $K_6$ topology, the additional symmetrical interaction within A-/B-sublattices extends the requirements to inequality in the internal and shell masses and to the internal acoustic anisotropy.

In the sequel, we describe the analytical model of metamaterials concerned and will conduct preliminary analyses on the properties of the acoustic system. The morphology of dispersion surfaces[†] ($\omega^2(\mathbf{k})$-surfaces), and the conditions of their contact for the case of the lattice in the absence of the internal structure are discussed. We present the results of our numerical analyses in the following way: the formation of band gaps in "symmetrical" cases (when the internal springs in both A- and B-sublattises determine the isotropic acoustic properties of the internal structures) is first analyzed, then further investigation is dedicated to the study of the formation of band gaps in the "partially symmetrical" system (when A-nodes have horizontal and B-nodes have vertical internal microstructures), finally the conditions of separation of the acoustic surfaces are obtained and discussed. It is shown that, for some acoustic surfaces, we will also obtain negative group velocity over a significant part of the Brillouin zone which makes presented metamaterial potentially beneficial in acoustic wave control, blocking low-frequency sound, and vibration shielding.

## 2. Equations of motion and dispersion

The elastic structures considered ($K_{3,3}$ and $K_6$[†]) comprise a pair of sublattices (A-sublattice and B-sublattice as depicted in Fig.1a) both with hexagonal symmetry. Topologies are depicted in Fig. 1b. The nodes that make up each sublattice are equivalent and have the same amplitude of vibration. As a reduction, considering each sublattice separately and neglecting interaction between them boils down the case to the problem $K_3$. The primitive Bravais cell contains two nodes. Thus, we have to write two pairs of equations of motion: for "red" and "blue" nodes taking into account the displacements of both external and internal masses. The external and internal masses in the A- and B- sublattices are denoted as $M_a, M_b, m_a, m_b$, respectively. The equations of motion for the two reference nodes A and B (See Fig. 1a) can be written as:

$$\begin{aligned} M_a \ddot{\mathbf{u}}_{ea} &= \mathbf{F}_{ea} - \mathbf{F}_{ia}, \\ m_a \ddot{\mathbf{u}}_{ia} &= \mathbf{F}_{ia}, \\ M_b \ddot{\mathbf{u}}_{eb} &= \mathbf{F}_{eb} - \mathbf{F}_{ib}, \\ m_b \ddot{\mathbf{u}}_{ib} &= \mathbf{F}_{ib}. \end{aligned} \quad (1)$$

where $\mathbf{u}_{ea}, \mathbf{u}_{eb}, \mathbf{u}_{ia}, \mathbf{u}_{ib}$ are the displacements of the reference node shells and the internal masses, respectively; $\mathbf{F}_{ia}, \mathbf{F}_{ib}$ are the forces exerted by the node shells on the internal masses, $\mathbf{F}_{ea} = \mathbf{F}_{e,aa} + \mathbf{F}_{e,ab}$, where $\mathbf{F}_{e,aa}$ is the force exerted by the six "red" nodes that surround the "red" central nodes A (See Fig. 2a), $\mathbf{F}_{e,ab}$ - the force exerted by the six "blue" nodes of the sublattice B (See Figs.1, 2b). Analogously for the reference node B: $\mathbf{F}_{eb} = \mathbf{F}_{e,bb} + \mathbf{F}_{e,ba}$.

---

[†] $K_n$ refers to the complete graph on n vertices and $K_{r,s}$ to the complete bipartite graph with r and s group vertex cardinalities.

The set of Eq.'s (1) relates to 20 external displacement vectors, but only two of them are independent: $\boldsymbol{u}_{ea}$ and $\boldsymbol{u}_{eb}$. Using Bloch's theorem the displacements of the six "red" nodes in Fig. 2a, $\boldsymbol{u}_{ea,j}$, for propagating plane waves can be presented as:

$$\boldsymbol{u}_{ea,j} = \boldsymbol{u}_{ea} \exp[i\sqrt{3}a(\boldsymbol{k} \cdot \boldsymbol{e}_j)],$$
$$\boldsymbol{e}_j = (cos\alpha_j, sin\alpha_j), \quad j = 1,2,\ldots,6. \quad \{\alpha_j\} = \left(\frac{\pi}{6}, \frac{\pi}{2}, \frac{5\pi}{6}, \frac{7\pi}{6}, \frac{3\pi}{2}, \frac{11\pi}{6}\right); \quad (2)$$

where $\boldsymbol{k}$ is the wave vector.

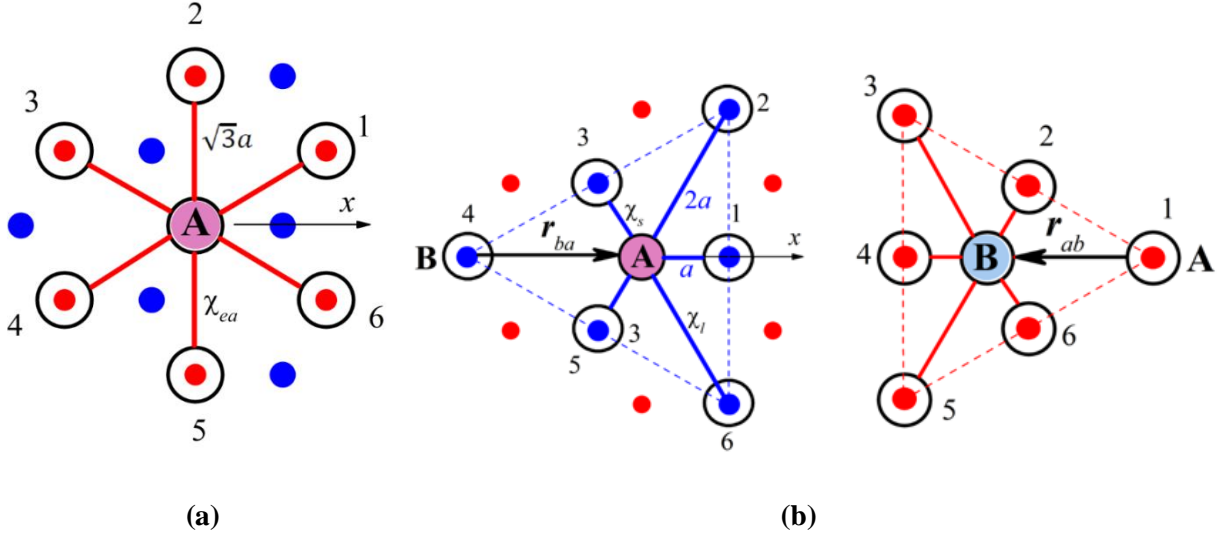

**(a)** **(b)**

**Fig. 2: a-** Interaction of the reference node A with neighbouring nodes of the same sublattice ("red"/A-sublattice). **b** – Interaction the reference nodes A and B with the nodes of the other sublattice (with B-and A-sublattice, respectively).

The displacements of the 6 "blue" nodes in Fig. 2b, $\boldsymbol{u}_{ea,j}$, are related to the displacement of the reference node B, $\boldsymbol{u}_{eb}$, as follow:

$$\boldsymbol{u}_{eb,j} = \boldsymbol{u}_{eb} \exp[i\boldsymbol{k} \cdot (\boldsymbol{r}_{ba} + d_j a \hat{\boldsymbol{e}}_j)]$$
$$\hat{\boldsymbol{e}}_j = (cos\beta_j, sin\beta_j), \quad \{\beta_j\} = \left(0, \frac{\pi}{3}, \frac{2\pi}{3}, \pi, \frac{4\pi}{3}, \frac{5\pi}{3}\right). \quad (3)$$
$$\boldsymbol{r}_{ba} = (2a, 0), \quad \{d_j\} = (1,2,1,2,1,2).$$

Let us write the set of equations for the displacement, $\boldsymbol{u}_{ea}$, of node A in detail.

$$\boldsymbol{F}_{e,aa} = \sum_{j=1}^{6} \boldsymbol{f}_{aa,j},$$

where $\boldsymbol{f}_{aa,j} = \chi_{ea} \boldsymbol{e}_j [\boldsymbol{e}_j \cdot (\boldsymbol{u}_{ea,j} - \boldsymbol{u}_{ea})]$ is the elastic force between two neighbouring nodes assuming linear elastic spring behaviour where $\chi_{ea}$ is the stiffness of the external springs in the A-sublattice.

Therefore:

$$\boldsymbol{F}_{e,aa} = \chi_{ea} \sum_{j=1}^{6} \boldsymbol{e}_j [\boldsymbol{e}_j \cdot \boldsymbol{u}_{ea} (\exp[i\sqrt{3}a(\boldsymbol{k} \cdot \boldsymbol{e}_j)] - 1)] \quad (4)$$

Then, $\boldsymbol{F}_{e,ab} = \sum_{j=1}^{6} \boldsymbol{f}_{ab,j}$, where $\boldsymbol{f}_{ab,j} = \chi_{ab,j}\hat{\boldsymbol{e}}_j[\hat{\boldsymbol{e}}_j \cdot (\boldsymbol{u}_{eb,j} - \boldsymbol{u}_{ea})]$ is the elastic force between the reference node A and the *j*-node of sublattice B (See Fig. 2b), and $\chi_{ab,j}$ is the stiffness of the corresponding spring. Finally, taking into account Eq. (3), we have:

$$\boldsymbol{F}_{e,ab} = \sum_{j=1}^{6} \chi_{ab,j}\hat{\boldsymbol{e}}_j[\hat{\boldsymbol{e}}_j \cdot (\boldsymbol{u}_{eb}\exp[i\boldsymbol{k} \cdot (\boldsymbol{r}_{ba} + d_j a \hat{\boldsymbol{e}}_j)] - \boldsymbol{u}_{ea})],$$
$$\{\chi_{ab,j}\} = (\chi_s, \chi_l, \chi_s, \chi_l, \chi_s, \chi_l), \tag{5}$$

where $\chi_s$ is the stiffness of short springs, and $\chi_l$ the stiffness of the long springs.
Analogously, for the reference node B we can write:

$$\boldsymbol{F}_{e,bb} = \chi_{eb}\sum_{j=1}^{6} \boldsymbol{e}_j[\boldsymbol{e}_j \cdot \boldsymbol{u}_{eb}(\exp[i\sqrt{3}a(\boldsymbol{k} \cdot \boldsymbol{e}_j)] - 1)],$$

$$\boldsymbol{F}_{e,ba} = \sum_{j=1}^{6} \chi_{ba,j}\hat{\boldsymbol{e}}_j[\hat{\boldsymbol{e}}_j \cdot (\boldsymbol{u}_{ea}\exp[i\boldsymbol{k} \cdot (\boldsymbol{r}_{ab} + l_j a \hat{\boldsymbol{e}}_j)] - \boldsymbol{u}_{eb})], \tag{6}$$

$$\boldsymbol{r}_{ab} = (-2a, 0), \quad \{\chi_{ba,j}\} = (\chi_l, \chi_s, \chi_l, \chi_s, \chi_l, \chi_s), \quad \{l_j\} = (2,1,2,1,2,1),$$

where $\chi_{eb}$ is the stiffness of the external springs in the B-sublattice.

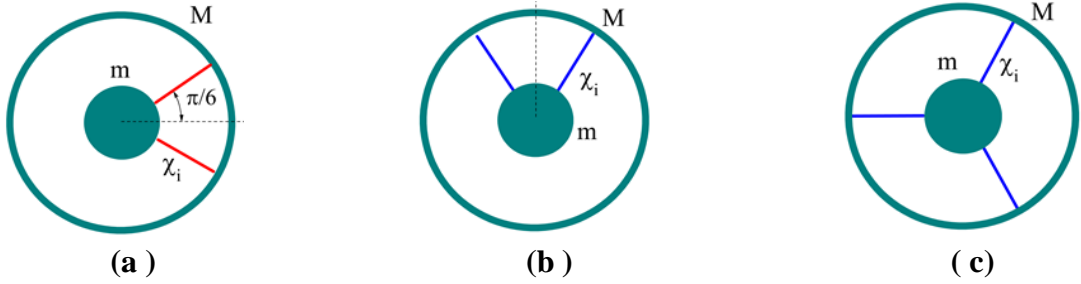

**Fig. 3.** Possible internal structures of the nodes. (a) anisotropic phononic microstructure internally stiffer in the horizontal direction, (b) anisotropic phononic microstructure internally stiffer in the vertical direction, (c) isotropic phononic microstructure

In the general case, the force exerted by external mass on the internal mass is equal to

$$\boldsymbol{F}_i = \chi_i \sum_n \boldsymbol{e}_n^{(i)}\left[\boldsymbol{e}_n^{(i)} \cdot (\boldsymbol{u}_e - \boldsymbol{u}_i)\right], \tag{7}$$

where $\boldsymbol{e}_n^{(i)} (n = 1,2..)$ are the unit vectors directed along the internal springs, $\boldsymbol{u}_e, \boldsymbol{u}_i$ – the displacement of the node shell and the internal mass, $\chi_i$ – the stiffness of the internal springs. In particular cases (See Fig. 3):

$$F_{ix} = \chi_i\tfrac{3}{2}u_{ex} - \chi_i\tfrac{3}{2}u_{ix}, \; F_{iy} = \chi_i\tfrac{1}{2}u_{ey} - \chi_i\tfrac{1}{2}u_{iy}, \text{ (case } a\text{)}$$
$$F_{ix} = \chi_i\tfrac{1}{2}u_{ex} - \chi_i\tfrac{1}{2}u_{ix}, \; F_{iy} = \chi_i\tfrac{3}{2}u_{ey} - \chi_i\tfrac{3}{2}u_{iy}, \text{ (case } b\text{)} \tag{8}$$
$$F_{ix} = \chi_i\tfrac{3}{2}u_{ex} - \chi_i\tfrac{3}{2}u_{ix}, \; F_{iy} = \chi_i\tfrac{3}{2}u_{ey} - \chi_i\tfrac{3}{2}u_{iy}, \text{ (case } c\text{)}$$

In the case *a* (*b*) an isolated node (without the external elastic forces, i.e. $M\mathbf{u}_e + m\mathbf{u}_i = 0$) has two eigen-frequencies of vibration:

$$\lambda_x^2 = \frac{3}{2}\frac{\chi_i}{m}\left(1 + \frac{m}{M}\right), \quad \lambda_y^2 = \frac{1}{2}\frac{\chi_i}{m}\left(1 + \frac{m}{M}\right), \tag{9}$$

for so called longitudinal (horizontal in Fig. 3*a*) and transverse (vertical) vibrations.
For the case of Fig. 3c we have:

$$\lambda_y^2 = \lambda_x^2 = \frac{3}{2}\frac{\chi_i}{m}\left(1 + \frac{m}{M}\right) = \lambda^2, \tag{10}$$

and the two branches of oscillations (BO) are degenerate.

Eventually, taking into account Eq.'s (2)-(8), the right part of the set of Eq.'s (1) can be represented as:

$$\mathbf{F} = \widehat{\mathbf{F}}(\mathbf{k}) \cdot \mathbf{U}^T,$$
$$\mathbf{U} = (u_{ea}^{(x)}, u_{ea}^{(y)}, u_{ia}^{(x)}, u_{ia}^{(y)}, u_{eb}^{(x)}, u_{eb}^{(y)}, u_{ib}^{(x)}, u_{ib}^{(y)}). \tag{11}$$
$$\mathbf{F}^T = (F_{ea}^{(x)} - F_{ia}^{(x)}, F_{ea}^{(y)} - F_{ia}^{(y)}, F_{ia}^{(x)}, F_{ia}^{(y)}, F_{eb}^{(x)} - F_{ib}^{(x)}, F_{eb}^{(y)} - F_{ib}^{(y)}, F_{ib}^{(x)}, F_{ib}^{(y)}).$$

For the harmonic vibrations, $\mathbf{U} \sim e^{i\omega t}$, the set of Eq.'s (1) can be rewritten as:

$$\mathbf{D}(\omega, \mathbf{k}) \cdot \mathbf{U}^T = 0,$$
$$D_{ij}(\omega, \mathbf{k}) = -\widehat{F}_{ij}(\mathbf{k})/M_i - \delta_{ij}\omega^2, \tag{12}$$
$$\mathbf{M} = (M_a, M_a, m_a, m_a, M_b, M_b, m_b, m_b)$$

The dimension of the dynamic matrix $\mathbf{D}(\omega, \mathbf{k})$ is $8 \times 8$ in this case. Eq. (12) leads to the eigenvalue problem which requires the singularity of the dynamic matrix (Eq. (13)) as follows:

$$|\mathbf{D}(\omega^2, \mathbf{k})| = 0 \tag{13}$$

Eq. (13) results in the desired dispersion surfaces $\omega_n^2 = \omega_n^2(\mathbf{k})$, $n = 1, 2, 3, \ldots, 8$.
The structure of the $\widehat{\mathbf{F}}$ matrix (elements marked X depict non-zero entries) is represented in Table 1.

| X | X | X | 0 | X | X | 0 | 0 |
|---|---|---|---|---|---|---|---|
| X | X | 0 | X | X | X | 0 | 0 |
| X | 0 | X | 0 | 0 | 0 | 0 | 0 |
| 0 | X | 0 | X | 0 | 0 | 0 | 0 |
| X | X | 0 | 0 | X | X | X | 0 |
| X | X | 0 | 0 | X | X | 0 | X |
| 0 | 0 | 0 | 0 | X | 0 | X | 0 |
| 0 | 0 | 0 | 0 | 0 | X | 0 | X |

**Table 1**. The structure of the $\widehat{\mathbf{F}}$ matrix. The upper left $4 \times 4$ square matrix corresponds to the interaction within the A (red)-sublattice, the lower right one corresponds to the interaction within the B (blue)-sublattice, and the other eight non-diagonal elements describe the interaction between A- and B-sublattices.

Below, the internal structure for both sublattices ("red" and "blue") is supposed to correspond to Fig. 6a; $\chi_{ia}$ – the stiffness of internal springs in A-sublattice, $\chi_{ib}$ – that in B-sublattice; $\widehat{\mathbf{k}} = a\mathbf{k}$.

$$\hat{F}_{11} = 3\chi_{ea}\left[\cos\left(\tfrac{3}{2}\hat{k}_x\right)\cos(\tfrac{\sqrt{3}}{2}\hat{k}_y) - 1\right] - 1.5(\chi_s + \chi_l) - 1.5\chi_{ia},$$
$$\hat{F}_{12} = F_{21} = -\sqrt{3}\chi_{ea}\sin\left(\tfrac{3}{2}\hat{k}_x\right)\sin(\tfrac{\sqrt{3}}{2}\hat{k}_y), \tag{14a}$$
$$\hat{F}_{22} = \chi_{ea}\left[2\cos(\sqrt{3}k_y) + \cos\left(\tfrac{3}{2}\hat{k}_x\right)\cos(\tfrac{\sqrt{3}}{2}\hat{k}_y) - 3\right] - 1.5(\chi_s + \chi_l) - 0.5\chi_{ia},$$
$$\hat{F}_{13} = \hat{F}_{31} = 1.5\chi_{ia}, \quad \hat{F}_{24} = \hat{F}_{42} = 0.5\chi_{ia}, \quad \hat{F}_{33} = -1.5\chi_{ia}, \quad \hat{F}_{44} = -0.5\chi_{ia},$$

$$\hat{F}_{15} = \sum \chi_{ab,j} \exp[i\hat{\boldsymbol{k}}(\boldsymbol{r}_{ba} + d_j \hat{\boldsymbol{e}}_j)] \cos^2\beta_j,$$
$$\hat{F}_{26} = \sum \chi_{ab,j} \exp[i\hat{\boldsymbol{k}}(\boldsymbol{r}_{ba} + d_j \hat{\boldsymbol{e}}_j)] \sin^2\beta_j$$
$$\hat{F}_{16} = F_{25} = \sum \chi_{ab,j} \exp[i\hat{\boldsymbol{k}}(\boldsymbol{r}_{ba} + d_j \hat{\boldsymbol{e}}_j)] \sin\beta_j\cos\beta_j, \tag{14b}$$
$$\hat{F}_{51} = \sum \chi_{ba,j} \exp[i\hat{\boldsymbol{k}}(\boldsymbol{r}_{ab} + l_j \hat{\boldsymbol{e}}_j)] \cos^2\beta_j = \hat{F}_{15}^*,$$
$$\hat{F}_{62} = \sum \chi_{ba,j} \exp[i\hat{\boldsymbol{k}}(\boldsymbol{r}_{ab} + l_j \hat{\boldsymbol{e}}_j)] \sin^2\beta_j = \hat{F}_{26}^*$$
$$\hat{F}_{61} = \hat{F}_{52} = \sum \chi_{ba,j} \exp[i\hat{\boldsymbol{k}}(\boldsymbol{r}_{ab} + l_j \hat{\boldsymbol{e}}_j)] \sin\beta_j\cos\beta_j = \hat{F}_{16}^* = \hat{F}_{25}^*,$$

$$\hat{F}_{55} = 3\chi_{eb}\left[\cos\left(\tfrac{3}{2}\hat{k}_x\right)\cos(\tfrac{\sqrt{3}}{2}\hat{k}_y) - 1\right] - 1.5(\chi_s + \chi_l) - 1.5\chi_{ib},$$
$$\hat{F}_{56} = \hat{F}_{65} = -\sqrt{3}\chi_{eb}\sin\left(\tfrac{3}{2}\hat{k}_x\right)\sin(\tfrac{\sqrt{3}}{2}\hat{k}_y), \tag{14c}$$
$$\hat{F}_{66} = \chi_{eb}\left[2\cos(\sqrt{3}k_y) + \cos\left(\tfrac{3}{2}\hat{k}_x\right)\cos(\tfrac{\sqrt{3}}{2}\hat{k}_y) - 3\right] - 1.5(\chi_s + \chi_l) - 0.5\chi_{ib},$$
$$\hat{F}_{57} = \hat{F}_{75} = 1.5\chi_{ib}, \quad \hat{F}_{68} = \hat{F}_{86} = 0.5\chi_{ib}, \hat{F}_{77} = -1.5\chi_{ib}, \hat{F}_{88} = -0.5\chi_{ib},$$

Above, we use the following identity laws:

$$\sum \exp(i\sqrt{3}\hat{\boldsymbol{k}}\boldsymbol{e}_j)\cos^2\alpha_j \equiv 3\cos\left(\tfrac{3}{2}\hat{k}_x\right)\cos(\tfrac{\sqrt{3}}{2}\hat{k}_y)$$
$$\sum \exp(i\sqrt{3}\hat{\boldsymbol{k}}\boldsymbol{e}_j)\sin\alpha_j\cos\alpha_j \equiv -\sqrt{3}\chi_{ea}\sin\left(\tfrac{3}{2}\hat{k}_x\right)\sin(\tfrac{\sqrt{3}}{2}\hat{k}_y)$$
$$\sum \exp(i\sqrt{3}\hat{\boldsymbol{k}}\boldsymbol{e}_j)\sin^2\alpha_j \equiv 2\cos(\sqrt{3}k_y) + \cos\left(\tfrac{3}{2}\hat{k}_x\right)\cos(\tfrac{\sqrt{3}}{2}\hat{k}_y)$$

To switch from the internal structure shown in Fig. 3a to that shown in Fig. 3b (for example, in the sublattice B), one must substitute the value $0.5\chi_{ib}$ for $1.5\chi_{ib}$ and inversely at the expression (14c). For symmetric configuration (Fig. 3c) only the substitution of $0.5\chi_{ib}$ for $1.5\chi_{ib}$ must be done.

## 3. Preliminary analysis of the properties of the acoustic system

The goal of preliminary analyses is the appropriate choice of parameters for further numerical investigation. In the most general case, Eq. (13) has 8 solutions, i.e. eight dispersion surfaces $\omega_n^2 = \omega_n^2(\boldsymbol{k})$, $n = 1,2,…,8$: two of which represent acoustic branches, $\omega_1^2(\boldsymbol{k})$, $\omega_2^2(\boldsymbol{k})$ [$\omega_1^2(\boldsymbol{k} = \boldsymbol{0}) = \omega_2^2(\boldsymbol{k} = \boldsymbol{0}) = 0$], and the other six the optical branches, $\omega_n^2(\boldsymbol{k})$, $n = 3,4,5,6,7,8$; $\omega_n^2(\boldsymbol{k} = \boldsymbol{0}) \neq 0$. The topology of these surfaces depend on at least eight dimensionless parameters (two parameters of the set of ten parameters $\{\chi_{ia}, \chi_{ib}, m_a, m_b, M_a, M_b, \chi_s, \chi_l, \chi_{ea}, \chi_{eb}\}$ can be used as units of mass and spring stiffness). In our case, we will operate with all of these ten parameters presented in some arbitrary units.

Fig. 4 shows an example of the solution to Eq. (13). Below, the dispersion surfaces $\omega^2(\hat{\boldsymbol{k}})$ will be presented in the rectangle $0 \leq k_x \leq 2\pi/3$, $0 \leq k_y \leq 4\pi/(3\sqrt{3})$, which include the right-upper quarter of the first Brillouin zone.

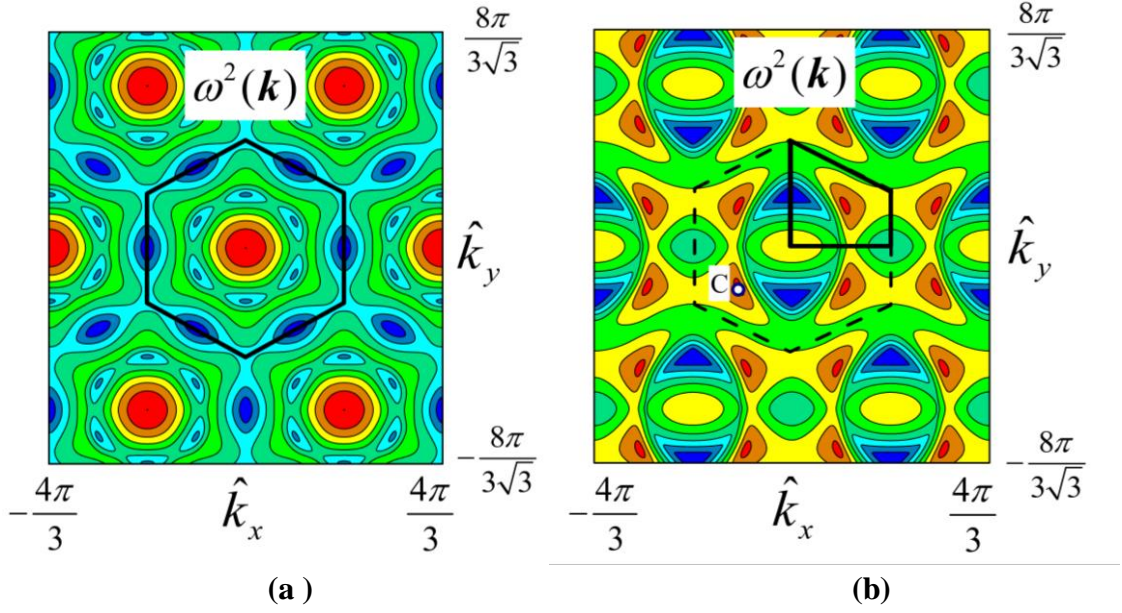

(a)                  (b)

**Fig.4**. An example of the $\omega^2(\hat{\boldsymbol{k}})$-distribution on the second from above optical frequency-surface, $m_b = m_a = 1$, $M_a = M_b = 2, \chi_s = \frac{4}{3}, \chi_l = 2/3$, $\chi_{ea} = \chi_{eb} = 0.3$. a - $\chi_{ib} = \chi_{ia} = 0$; the solid line bounds the first Brillouin zone (BZ). b - $\chi_{ib} = \chi_{ia} = 1$. The internal springs destroy the hexagonal symmetry of the lattice. So, it is not enough to analyse the structure of the dispersion surfaces on $1/6^{\text{th}}$ part of the BZ, and one must consider a quarter of BZ instead. Besides, the analysis of $\omega^2(\boldsymbol{k})$, only on the closed contour shown in Fig. 4b is not sufficient either, due to possible existence of position of extrema of that function inside the quarter of BZ (like the point C in Fig. 4b).

As the first step towards a comprehensive analysis of the system, we need to know the frequency characteristics of the acoustic systems under consideration at simple extreme cases.

(i)      The eigen-frequencies of the isolated nodes of the A-and B-sublattices (free internal vibrations, or formally the case of zero extrenal stiffness $\chi_{ea} = \chi_{eb} = \chi_s = \chi_l = 0$) are equal to:

$$\lambda^2_{x,a,b} = \frac{3}{2}\frac{\chi_{i,a,b}}{m_{a,b}}\left(1+\frac{m_{a,b}}{M_{a,b}}\right), \quad \lambda^2_{y,a,b} = \frac{1}{2}\frac{\chi_{i,a,b}}{m_{a,b}}\left(1+\frac{m_{a,b}}{M_{a,b}}\right). \tag{15}$$

A very important remark must be made at this juncture. If

$$\lambda^2_{xa} = \lambda^2_{xb} = \lambda^2_x, \quad \lambda^2_{ya} = \lambda^2_{yb} = \lambda^2_y, \tag{16}$$

then Eq. (13), $|D(\omega^2 = \lambda^2_x, \boldsymbol{k}=\boldsymbol{0})| = 0, |D(\omega^2 = \lambda^2_y, \boldsymbol{k}=\boldsymbol{0})| = 0$ holds for arbitrary values of all other parameters. In other words, in the case of the equalities of Eq. (16) the frequencies of the two optical frequency-surfaces, $i,j$, are equal to:

$$\omega_i^2(k=0) = \lambda_x^2 \text{ and } \omega_j^2(k=0) = \lambda_y^2 \tag{17}$$

at the centre of the BZ.

(ii) If the nodes of the A-and B-sublattices do not possess an internal structure (formally, if $\chi_{i,a,b} = 0$), there are only two branches of vibrations in Eq. (13) that have the same frequencies at $\boldsymbol{k} = 0$ (on the condition that $\omega^2(k = 0) \neq 0$).

$$\widehat{\omega}^2_{opt,1,2}(\boldsymbol{k} = 0, \chi_{ia,b} = 0) = \frac{3}{2}(\chi_s + \chi_l)\left(\frac{1}{M_a} + \frac{1}{M_b}\right), \tag{18}$$

These obviously do not depend on the stiffnesses $\chi_{ea}, \chi_{eb}$ and both correspond to "optical phonons" when A- and B-sublattice oscillate anti-phase to each other along X- and Y-axis:

$$u_{ea,x} = u^{(0)}_{ea,x}\exp(i\widehat{\omega}_{opt,1}t),\ u_{eb,x} = u^{(0)}_{eb,x}\exp(i\widehat{\omega}_{opt,1}t - i\pi),\ u^{(0)}_{ea,y} = u^{(0)}_{eb,y} = 0,$$
$$u_{ea,y} = u^{(0)}_{ea,y}\exp(i\widehat{\omega}_{opt,2}t),\ u_{eb,x} = u^{(0)}_{eb,y}\exp(i\widehat{\omega}_{opt,2}t - i\pi),\ u^{(0)}_{ea,x} = u^{(0)}_{eb,x} = 0. \tag{19}$$

where $u_{ea,x}, u_{ea,y}, u_{eb,x}, u_{eb,x}$ are the displacements of the reference nodes A and B.
There are two the acoustic branches in this case as well ($\omega^2(k=0) = 0$).

(iii) It is useful to know the characteristic frequencies of acoustic vibrations for the case of $\chi_{i,a,b} = 0$, $\chi_{s,l} = 0$. These frequencies (in the units of $\chi_{ea}/M_a$ and $\chi_{eb}/M_b$ for A-and B-sublattice that are independent systems in this case) are equal to:

$$\widehat{\omega}^2_{ac,1,2}\big|_{A,B} = (\chi_{ea,b}/M_{a,b}) \times f(\widehat{k}_x, \widehat{k}_y), \tag{20a}$$

$$f(\widehat{k}_x, \widehat{k}_y) = 3 - 2\cos\left(\frac{3}{2}\widehat{k}_x\right)\cos\left(\frac{\sqrt{3}}{2}\widehat{k}_y\right) - \cos(\sqrt{3}\widehat{k}_y) \pm$$
$$\pm \sqrt{\left[\cos\left(\frac{3}{2}\widehat{k}_x\right)\cos\left(\frac{\sqrt{3}}{2}\widehat{k}_y\right) - \cos(\sqrt{3}\widehat{k}_y)\right]^2 - 3\sin^2\left(\frac{3}{2}\widehat{k}_x\right)\sin^2(\frac{\sqrt{3}}{2}\widehat{k}_y)}. \tag{20b}$$

The upper frequency surfaces attain their maximum on the boundary of the BZ as:

$$\widehat{\omega}^2_{ac,2}\left(k_x = \frac{2\pi}{3}, k_y = 0\right)\bigg|_{A,B} = 6 \times (\chi_{ea,b}/M_{a,b}). \tag{21}$$

## 4. Morphology of the $\omega^2(k)$-surfaces for simple cases ($\chi_{ia,b} = 0$).

In order to reduce the number of parameters in our numerical experiments we suppose $1/M_a + 1/M_b = 1$, and $\chi_s + \chi_l = 1$, and that the stiffnesses of the springs are inversely proportional to the length of the springs (a simple rod analogy). So, $\chi_s/\chi_l = 2$, $\chi_{ea} = \chi_{eb} = \chi_s/\sqrt{3}$ in the problem K$_6$.

In all the cases shown in Fig. 5 the optical frequency-surfaces at the center of the BZ pass through the same point i.e. the frequencies are equal to the same value (See Eq. (18)).

$$\widehat{\omega}^2_{opt,1,2}(\boldsymbol{k} = 0, \chi_{ia,b} = 0) = 3/2, \tag{22}$$

This value is the maximum frequency over the BZ in the case of the K$_{3,3}$-system (Fig.s 5a,b). In the case of the K$_6$-system, the additional springs that operate within the A- and B-nodes ($\chi_{ea} \neq 0$, $\chi_{eb} \neq 0$) excite the oscillation, as described by Eq. (20). Combination of these oscillations with other characteristic frequencies of the entire system results in increasing the frequencies at the periphery of the Brillouin zone: for the optical branches $\omega^2(\boldsymbol{k}=\boldsymbol{0}) < \omega^2(\boldsymbol{k} \neq \boldsymbol{0})$ (Fig.s 5c,d).

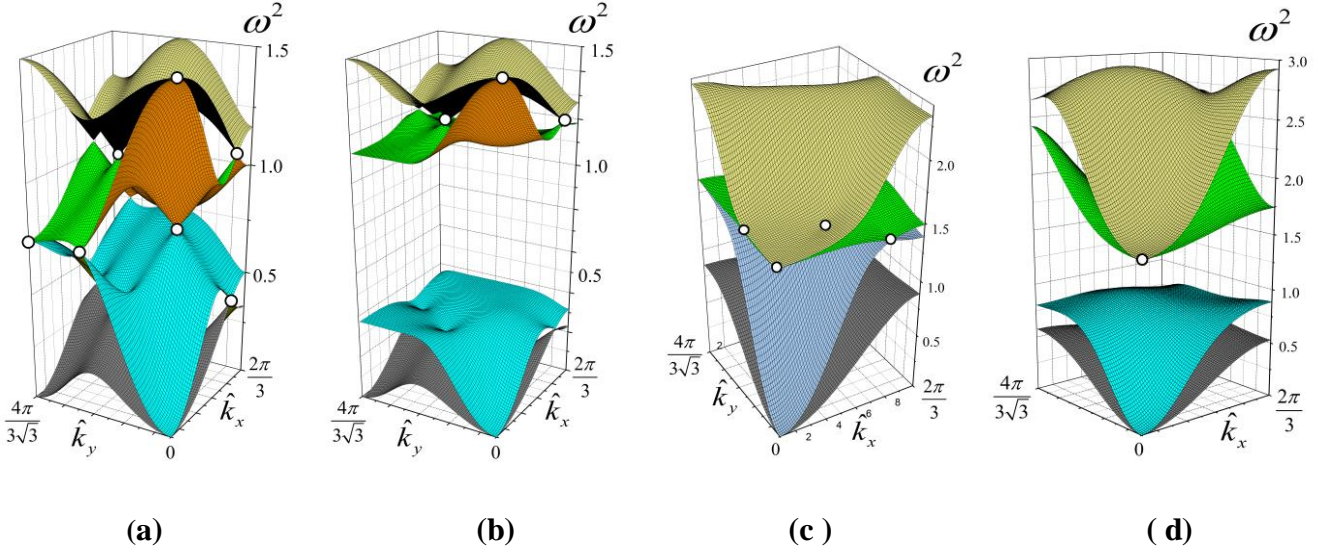

(a) (b) (c) (d)

**Fig. 5.** The frequency surfaces in the simplest case when $\chi_{ia} = \chi_{ib} = 0$. The stiffnesses $\chi_s, \chi_l$ are inversely proportional to the length of the springs: $\chi_s = 2/3, \chi_l = 1/3$.
**a, b** – the $K_{3,3}$-problem ($\chi_{ea} = \chi_{eb} = 0$): **a**- $M_a = M_b = 2$; **b** - $M_a = 4, M_b = 4/3$. Points of tangential contact on the dispersion surfaces are marked with white circles.
**c, d** – the $K_6$-problem ($\chi_{ea} = \chi_{eb} = \chi_s/\sqrt{3} = 2/3\sqrt{3}$.): **c**- $M_a = M_b = 2$; **d**- $M_a = 4, M_b = 4/3$.
In all of the presented cases $\chi_s + \chi_l = 1, 1/M_a + 1/M_b = 1$, and $\widehat{\omega}^2_{opt,1,2}(\boldsymbol{k}=\boldsymbol{0}, \chi_{ia,b}=0) = 3/2$ (See Eq. (19)).
The greater the ratio of masses, the wider is the frequency gap between the upper acoustic and lower optic surfaces.

If the masses of A- and B-nodes are equal, the frequency-surfaces contact each other at a number of points (Fig.s 5a,c). At these points repulsion between neighbouring $n$ and $j$ – "energy level" ($\omega_n^2(\boldsymbol{k}) = \omega_j^2(\boldsymbol{k})$, $\boldsymbol{k}_n = \boldsymbol{k}_j$) does not occur, if the corresponding oscillators are in different states. In other words,

$$\boldsymbol{u}_{a,n} \perp \boldsymbol{u}_{a,j}, \boldsymbol{u}_{b,n} \perp \boldsymbol{u}_{b,j}. \quad (23)$$

As the unit cell contains two nodes, the displacements of the nodes may be circularly polarised, which corresponds to the case of complex eigenvectors as:

$$\boldsymbol{U}_n(u^{(x)}_{ea,n}, u^{(y)}_{ea,n}, u^{(x)}_{eb,n}, u^{(y)}_{eb,n}) \text{ and } \boldsymbol{U}_j(u^{(x)}_{ea,j}, u^{(y)}_{ea,j}, u^{(x)}_{eb,j}, u^{(y)}_{eb,j}).$$

So, the conditions of Eq. (23), could be formally presented as:

$$\begin{aligned} u^{(x)}_{ea,n} \cdot u^{(x)}_{ea,j} + u^{(y)}_{ea,n} \cdot u^{(y)}_{ea,j} &= 0, \\ u^{(x)}_{eb,n} \cdot u^{(x)}_{eb,j} + u^{(y)}_{eb,n} \cdot u^{(y)}_{eb,j} &= 0, \end{aligned} \quad (24)$$

which are stronger than the orthogonality condition $U_n^* \cdot U_j^T = 0$. These relations (24) are satisfied at the points of tangential contact with high accuracy in our numerical experiments.

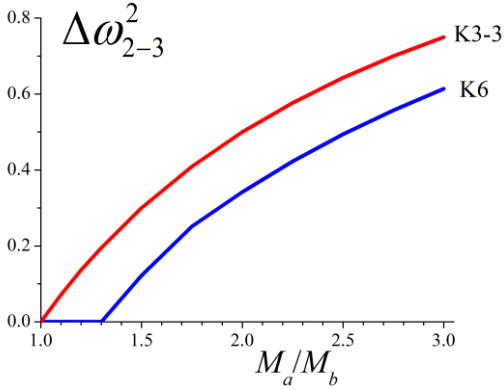

**Fig. 6.** Dependency of the band gap between the upper acoustic, $\omega_2^2(\boldsymbol{k})$, and the lower optical, $\omega_3^2(\boldsymbol{k})$, surfaces on the mass ratio $M_a/M_b$ for the fixed value of $1/M_a + 1/M_b = 1$ and $\chi_s = 2/3$, $\chi_l = 1/3$, which conserve the optical frequency at the center of the BZ: $\widehat{\omega}_{opt,1,2}^2(\boldsymbol{k}=0) = \frac{3}{2}$ ($\chi_{ia,b} = 0$, no mass-in-mass structure in the nodes). For the $K_6$-configuration $\chi_{ea} = \chi_{eb} = 2/3\sqrt{3}$.

The inequality of the masses of A-and B-nodes destroys such specific condition (24) for the acoustic and optical modes (Fig.'s 5b,d). It looks like the lower optical surface repels the acoustic surfaces lowering them down. This effect can be envisaged as the classical analog of the quantum mechanical 'level repulsion' if the coupling appears between these levels. The greater the ratio of masses the broader is the frequency gap, $\Delta\omega_{2-3}^2$, between these modes (See Fig. 6 ). In the $K_6$-system the ratio $M_a/M_b$ must exceed some threshold for the frequency band gap to arise. This effect is conditioned by the additional acoustic waves (20) when $\chi_{ea}, \chi_{eb} \neq 0$.

## 5. The effects of mass-in-mass structure

As one can see, the inequality $M_a \neq M_b$ does not split the two "original" optical modes (See Fig.'s 5b,d) according to Eq. (18) if $1/M_a + 1/M_b = const$. The splitting can be realised due to exciting the internal vibrations in A-and-B-sublattices.

Let us qualitatively analyse the simplest situation at $\boldsymbol{k} = 0$ when $M_a = M_b = 2$ (as in Fig.'s 5a,c), the internal masses $m_a = m_b$, and the stiffnesses $\chi_{ia} = \chi_{ib}$ are variable but the frequencies $\lambda_{xa}^2 = \lambda_{xb}^2 = \lambda_x^2$ and $\lambda_{ya}^2 = \lambda_{yb}^2 = \lambda_y^2$ (See Eq.'s (15),(16)) stay fixed: $\chi_i \sim mM/(m+M)$ - indices $a$ and $b$ are ignored here. As above, in this case there are certainly two the new optic dispersion surfaces with the frequencies of $\lambda_x^2$ and $\lambda_y^2$ at $k = 0$. The external force field (external springs) exert on the node shells, which interact with internal masses. This interaction of external and internal vibrations generates four additional optical frequency-surfaces, which can be presented as follows.

If the condition of Eq. (16) is held, then three possible scenarios for interaction at $\boldsymbol{k} = 0$ can be realised.

(i) The external elastic springs vanish i.e. external forces are zero. In other words, two superlattices (the first includes both the A-and B- external masses and the second includes the A-and B-internal masses) vibrate in antiphase without deformation of the external springs ($\boldsymbol{u}_{ea} = \boldsymbol{u}_{eb}$). The displacements of the external and internal masses satisfy the ratios that correspond to the free internal vibrations: $m_a\boldsymbol{u}_{ia} + M_a\boldsymbol{u}_{ea} = \boldsymbol{0}$, and $m_b\boldsymbol{u}_{ib} + M_b\boldsymbol{u}_{eb} = \boldsymbol{0}$. Formally, the frequencies of the internal vibrations in nodes of the A- and-B-sublattices can be re-written (See Eq.'s (9)) as $\lambda_x^2 = \frac{3}{2}\frac{\chi_i}{m}(1 + m/\widehat{M}) -$ (indices $a$ and $b$ ignored here), $\lambda_y^2 = \frac{1}{2}\frac{\chi_i}{m}(1 + m/\widehat{M})$, where the effective mass $\widehat{M} = -m u_i/u_e$ is equal to the real mass, $M$, of the node shell for the free internal vibration (See Fig. 7; $\omega_5^2(\mathbf{k}=\mathbf{0}) = \lambda_y^2$, $\omega_7^2(\mathbf{k}=\mathbf{0}) = \lambda_x^2$).

(ii) The external springs are present i.e. the external vibratory force field is not equal to $\mathbf{0}$. Then internal vibrations can be brought to the "free" internal vibrations with the effective mass of the shells. The self-consistency of the external and internal forces results in only two possible effective masses, $\widehat{M}^{(+)}, \widehat{M}^{(-)}$, for both vibration polarisations (along X- and Y-axis: $\widehat{M}_x^{(+)}, \widehat{M}_x^{(-)}$ and $\widehat{M}_y^{(+)}, \widehat{M}_y^{(-)}$, respectively). A peculiarity is that for any set of parameters, which meet the conditions of Eq. (16), $0 < \widehat{M}_x^{(+)}, \widehat{M}_y^{(+)} < M$. This means that the internal masses and the node shells oscillate anti-phase ($\mathbf{u}_i \cdot \mathbf{u}_e < 0$), but the displacements of the shells are greater than by the real free internal vibrations. Thus, the effective stiffnesses of the internal springs, $\hat{\chi}_i = \chi_i(1 + m/\widehat{M})$, increase. This way two optical frequencies arise (See. Fig. 7):

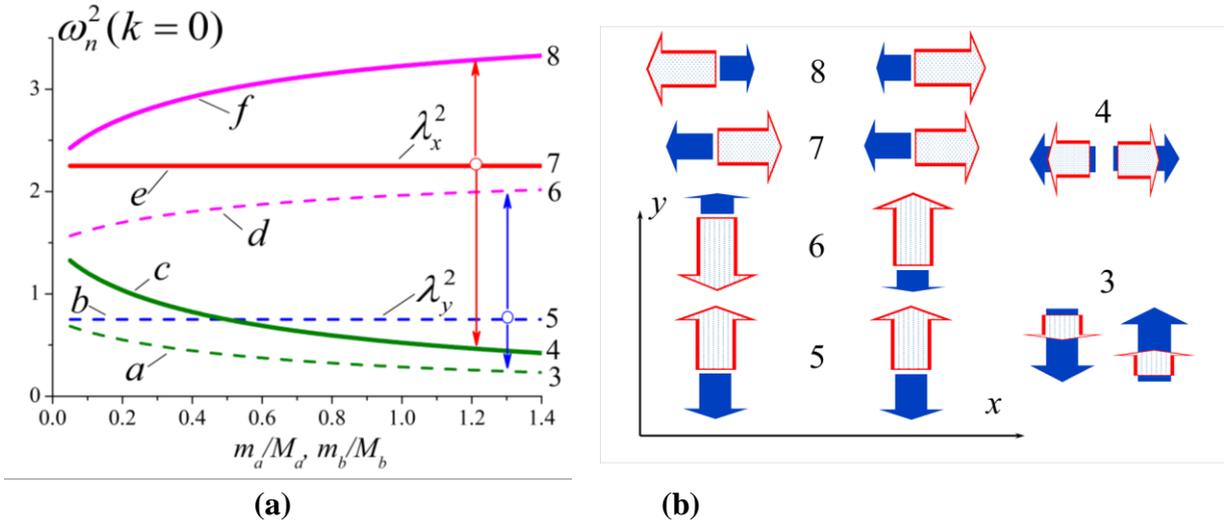

(a) (b)

**Fig. 7.** The scenarios of formation of the optics modes at the center of the BZ ($\mathbf{k}=\mathbf{0}$). $\chi_s = 2/3$, $\chi_l = 1/3$, $M_a = M_b = 2$ (the same as in Fig. 5, 6). Changing the internal masses ($m_a = m_b$) are followed by changing the stiffnesses of the internal springs so that $\lambda_x^2 = 2.25$, $\lambda_y^2 = 0.75$; $(\lambda_x^2 + \lambda_y^2)/2 = \widehat{\omega}_{opt,1,2}^2(\mathbf{k}=\mathbf{0}, \chi_{ia,b} = 0)$. $a - \omega_8^2 = \omega^2(\widehat{M}_x^{(+)}, \mathbf{k}=\mathbf{0})$, $\omega_7^2(\mathbf{k}=\mathbf{0}) = \lambda_x^2$, $\omega_6^2 = \omega^2(\widehat{M}_y^{(+)}, \mathbf{k}=\mathbf{0})$, $\omega_5^2(\mathbf{k}=\mathbf{0}) = \lambda_y^2$, $\omega_4^2 = \omega^2(\widehat{M}_x^{(-)}, \mathbf{k}=\mathbf{0})$, $\omega_3^2 = \omega^2(\widehat{M}_y^{(-)}, \mathbf{k}=\mathbf{0})$.

$b$ – graphical presentation of the corresponding displacements $\mathbf{u}_e(\omega_n^2(\mathbf{k}=\mathbf{0}))$ - red bounded arrows, and $\mathbf{u}_i(\omega_n^2(\mathbf{k}=\mathbf{0}))$ – dark blue arrows; $\frac{m}{M} = 1.4$. At left for B-nodes, and at right for the A-nodes.

$$\omega_6^2\left(\widehat{M}_y^{(+)}, k=0\right) = \frac{1}{2}\frac{\chi_i}{m}\left(1 + m/\widehat{M}_y^{(+)}\right) > \lambda_y^2,$$

$$\omega_8^2\left(\widehat{M}_x^{(+)}, k=0\right) = \frac{3}{2}\frac{\chi_i}{m}\left(1 + m/\widehat{M}_x^{(+)}\right) > \lambda_x^2.$$

(25)

(iii) In-phase oscillations of the internal and external masses ($\boldsymbol{u}_i \cdot \boldsymbol{u}_e > 0$) corresponds to the negative effective masses $\widehat{M}_x^{(-)}, \widehat{M}_y^{(-)}$ for the node shells ($\widehat{M} = -m\frac{\boldsymbol{u}_i}{\boldsymbol{u}_e} < 0$). So, for another pair of the optical frequencies (See Fig. 7) the following inequalities take place:

$$\boldsymbol{u}_i \cdot \boldsymbol{u}_e > 0, \quad \omega_4^2\left(\widehat{M}_x^{(-)}, k = 0\right) = \frac{3}{2}\frac{\chi_i}{m}\left(1 + m/\widehat{M}_x^{(-)}\right) < \lambda_x^2,$$

(26)

$$\omega_3^2\left(\widehat{M}_y^{(-)}, k = 0\right) = \frac{1}{2}\frac{\chi_i}{m}\left(1 + m/\widehat{M}_y^{(-)}\right) < \lambda_y^2.$$

It is to be noted that we have only discussed the scenarios occurring at the center of the BZ, which do not depend on the stiffnesses $\chi_{ea}, \chi_{eb}$ and are also valid in the investigation of the acoustic properties of the K$_6$ systems. (Formally, the summands in the matrix $D(\omega^2, \boldsymbol{k})$ that include the parameters $\chi_{ea}, \chi_{eb}$ vanish at $\boldsymbol{k} = \boldsymbol{0}$).

If the internal structure is isotropic (See. Fig. 3c) then $\lambda_y^2 = \lambda_x^2 \equiv \lambda^2$, and in Fig. 7a the line $b$ moves up to coincide with the line $e$. Correspondingly, the curve $a$ coincides with the curve $c$, and the curve $d$ with the curve $f$. Thus, the solid lines $c, e, f$ in the Fig. 7 display the typical solutions of the Eq. (13) when the internal frequency $\lambda^2$ stays invariable on changing the masses, $m_a, m_b$. Physically, this means that for any $\lambda^2$ the six optical surfaces merge into the three pairs of those at the center of the BZ. In other words, the three twice-degenerate vibration levels form at $\boldsymbol{k} = \boldsymbol{0}$.

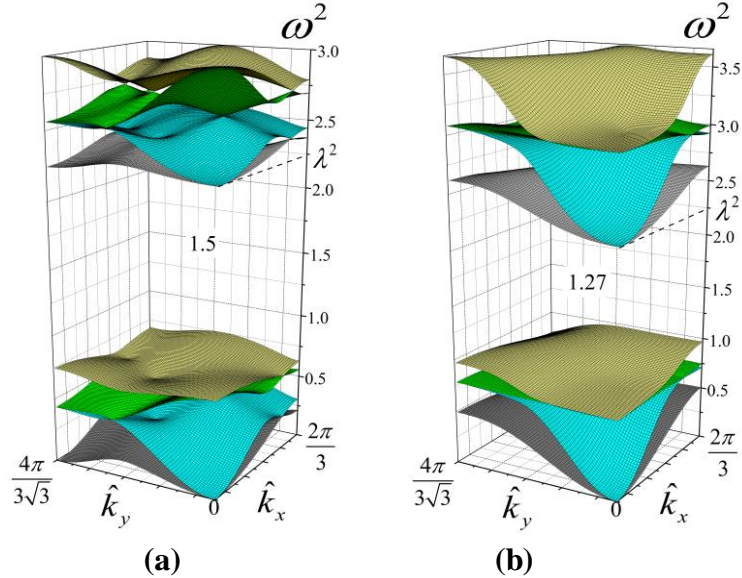

**Fig. 8.** The gap band formation due to being the isotropic internal structures in the A- and B- nodes (See Fig. 3c). $\chi_s = 2/3, \chi_l = 1/3, M_a = M_b = 2, m_a = m_b = 1, \chi_{ia} = \chi_{ib} = 1$ ($\lambda^2 = 2.25$). The gap bands, $\Delta\omega_{4-5}^2$, between the neighbor optical surfaces are shown. $a -$ The result for the K3-3 system. $b$- The frequency surfaces in case of K6 system ($\chi_{ea} = \chi_{eb} = \chi_s/\sqrt{3}$ ; as mentioned above, the stiffnesses of the external springs are inversely proportional to their length). In the both cases, on the two upper optical surfaces $\omega_{7,8}^2(\boldsymbol{k} = \boldsymbol{0}) = 3$, on the two lower optical surfaces $\omega_{3,4}^2(\boldsymbol{k} = \boldsymbol{0}) = 0.75, \omega_{5,6}^2(\boldsymbol{k} = \boldsymbol{0}) = \lambda^2$.

In Fig. 7, one can see that the single frequency-surfaces (or pairs of them) may be formed mainly at $m/M \gtrsim 1$. This ratio provides the effective coupling between the external and internal vibrations in formation of the general vibratory force field as a result of that $u_e \gtrsim u_i$ through the free internal vibrations, i.e. when the free internal vibrations "are not closed" inside of the node shells.

## 6. Results of the numerical simulations

As mentioned above, the gaps may be formed even when the nodes of A- and B-sublattices are absolutely identical. For the isotropic internal structures, at $k = 0$ (Fig. 7) the frequency band between the lower pair of the optical surfaces (See the curve $c$) and the middle pair (See the line $e$) is broad enough so that these pair do not contact over the entire BZ. The relatively narrow initial gap (the gap between the line $e$ and the curve $f$) may be eliminated at $k > 0$ (See. Fig. 8). In the case of the $K_6$-system, the additional acoustic modes of Eq. (20) broaden the frequency bands of the optical vibrations and the gap, $\Delta\omega^2_{4-5}$, is narrower than in case of the $K_{3,3}$-system.

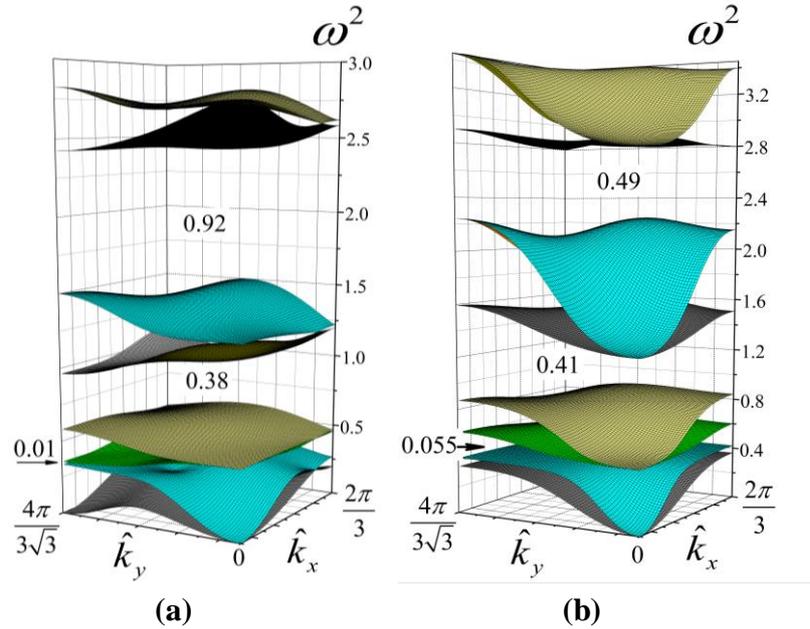

**Fig. 9.** Gap formation in the partially-symmetrical heterogeneous systems: the "horizontal" internal structure in A-nodes (Fig. 3a) is combined with the "vertical" internal structure in B-nodes (Fig. 3b).The set of the parameters used is shown in Fig. 8. **a** – the frequency surfaces for the $K_{3,3}$ system, **b** – the result for the $K_6$ system. In both the cases $\omega^2_{3,4}(k = 0) \approx 0.48$, $\omega^2_{5,6}(k = 0) \approx 1.27$, $\omega^2_{7,8}(k = 0) \approx 2.74$.

In both cases presented in Fig. 8, there are no single pairs of optical surfaces, in spite of there being an initial gap of size $\Delta\omega^2_{6-7}(k = 0) = \omega^2_7(k = 0) - \omega^2_6(k = 0) = 3 - \lambda^2 = 0.75$. At the center of the BZ on each surface pair $(\omega^2_n(k) \cup \omega^2_{n+1}(k), n = 3,5,7)$, the external/internal masses vibrate in mutually perpendicular directions, and these vibrations are linearly polarised. Outside of the center, the trajectories become intricate, however; points on the neighbouring surfaces can be found at which these trajectories are "perpendicular", too (See Fig. 8a, for instance). To remove the contacts between the frequency-surfaces we have to minimise the probability of formation of such points in which the Eq.'s (24) (the conditions of "perpendicularity" of Eq. (23)) are satisfied.

As a first approach to solving the problem we consider a "partially symmetrical" system. As above, the A- and B- nodes are absolutely identical and both sublattices possess anisotropic internal structures (See Fig. 3a,b). Nevertheless, the internal spring orientations are different: in the A-nodes – horizontal (Fig. 3a), in the B-nodes – vertical (Fig. 3b); i.e., the B-nodes are rotated relative to A-nodes through an angle of $\pi/2$. Thus, the directions along X- and Y-axis become in a certain sense equivalent (partly-symmetrical) i.e.:

$$\lambda_{xa}^2 + \lambda_{xb}^2 = \lambda_{ya}^2 + \lambda_{yb}^2. \tag{27}$$

This partial symmetry, just as in the case of the isotropic symmetry above, causes the formation of three twice-degenerate vibration levels at $\boldsymbol{k} = \boldsymbol{0}$, and at the same time reduces the symmetry properties of the entire system that results in eliminating the tangential points of contact between neighbouring pairs of frequency-surfaces (See. Fig. 9).

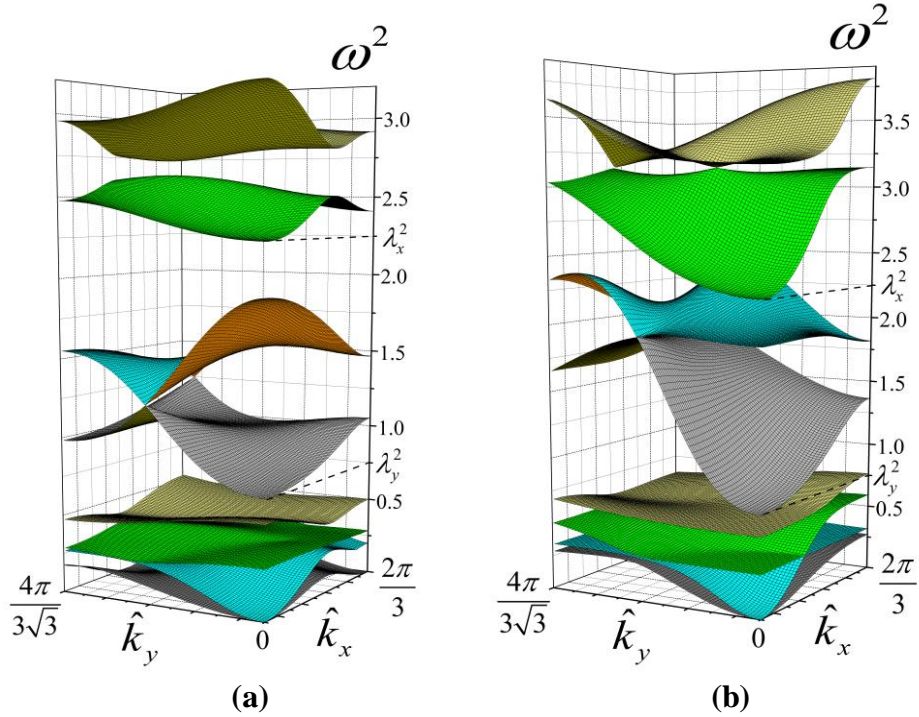

**Fig. 10.** Results for $K_{3,3}$-system (a) and $K_6$-system (b) composed of the identical A- and B-sublattices, but with internal vibrations of the entire system as anisotropic. $M_a = M_b = 2$, $m_a = m_b = 1.6$, $\chi_{ia}=\chi_{ib}=4/3$, $\chi_s = 2/3$, $\chi_l = 1/3$, $\chi_{ea} = \chi_{eb} = \chi_s/\sqrt{3}$ (in the $K_6$-system, only). Orientation of the internal springs corresponds to Fig. 3a: $\lambda_{xa}^2 = \lambda_{xb}^2 = 2.25$, and $\lambda_{ya}^2 = \lambda_{yb}^2 = 0.75$. The frequencies $\omega_n^2(k = 0)$, $n = 3,4,5,...,8$ correspond to those shown in Fig. 7a at $m_{a,b}/M_{a,b} = 0.8$.

It must be noted that in both cases ($K_{3,3}$ and $K_6$) the optical frequencies at $k = 0$ are the same. However, the total band gap, $\Delta\Omega^2 = \Delta\omega_{2-3}^2 + \Delta\omega_{4-5}^2 + \Delta\omega_{6-7}^2$ (as depicted in Fig. 9) in the $K_6$-system is lower than in the $K_{3,3}$-system due to the additional acoustic modes (See Eq. (20)), which broaden each of the optical frequency bands. On the other hand, these additional phonons (modes) intensify the "repulsion" between the neighbouring dispersion surfaces (compare Fig. 9a and Fig. 9b), an effect which appears most conspicuously between the upper acoustic mode, $\omega_2^2(\boldsymbol{k})$, and the lower optical mode, $\omega_3^2(\boldsymbol{k})$: the gap $\Delta\omega_{2-3}^2$ in Fig. 9b ($K_6$-system) is much broader than that in Fig. 9a ($K_{3,3}$-system).

As shown by the numerical simulations conducted, the optimal set of parameters for gap formation corresponds to the ratio:

$$(\lambda_{xa}^2 + \lambda_{xb}^2)/2 = (\lambda_{ya}^2 + \lambda_{yb}^2)/2 \approx \widehat{\omega}_{opt,1,2}^2(\boldsymbol{k} = \boldsymbol{0}, \chi_{ia,b} = 0). \tag{28}$$

Otherwise, the lower gap, $\Delta\omega_{2-3}^2$, may disappear. As the internal masses increase, the gap size $\Delta\omega_{2-3}^2$ decreases. Qualitatively, this can be interpreted on the basis of the results shown in Fig. 7a: the frequencies $\omega_3^2(k=0)$ and $\omega_4^2(k=0)$ abate for higher internal masses. Correspondingly, the gaps $\Delta\omega_{4-5}^2$ and $\Delta\omega_{6-7}^2$ slightly broaden. At the low value of mass ratios $\frac{m_a}{M_a} = \frac{m_b}{M_b} \ll 1$ the gap, $\Delta\omega_{2-3}^2$, disappears and the maximal value of $\Delta\omega_{2-3}^2$ corresponds to the ratio $\frac{m_{a,b}}{M_{a,b}} \sim 0.4 - 0.6$.

Thus, by constructing the elastic system consisting of absolutely identical nodes we can form the gaps between the *pairs* of the acoustic and optical modes. On constricting the symmetry properties of the entire system (from the isotropic internal structure to the partially symmetrical case), the coupling between the neighbouring vibration levels intensifies, which results in slitting some frequency surfaces not only at the center but over the whole BZ (Fig.'s 8, 9).

Next step would be to form the *single* optical surface which requires further reduction in symmetry (for some conditions this symmetry not to be realised at all). As one can see in Fig. 7a, there is no difficulty in splitting the optical modes at $k = 0$ using the anisotropic internal vibrations in the entire system when the orientation of the internal springs are the same in both A-and B-sublattices (for instance, all horizontal –See. Fig. 3a). But such anisotropy is not adequate to solve the problem fully (See. Fig. 10) since frequency-surfaces overlap at the periphery of the BZ.

Therefore, the next approach is to use different internal masses in the A- and B-sublattices by satisfying the equalities (16): $\lambda_{xa}^2 = \lambda_{xb}^2$, $\lambda_{ya}^2 = \lambda_{yb}^2$. In the sequel, we present the results obtained for the $K_6$- and $K_{3,3}$-systems separately, starting with the former.

**6.1.   Dispersion surfaces of the $K_6$-system**. As shown by the result of numerical experiments conducted, there is only one method to form the single optical surfaces in $K_6$-system (See Fig. 11) i.e. by using substantially different internal masses at A- and B-nodes (on the assumption that $M_a = M_b$). But for all that, the internal frequencies are:

$$\lambda_x^2, \lambda_y^2 > \widehat{\omega}_{opt,1,2}^2(\boldsymbol{k} = \boldsymbol{0}, \chi_{ia,b} = 0), \tag{29}$$

and one of the internal masses essentially exceeds the corresponding external mass. The physical meaning of the condition (29) is obvious. As follows from Eq. (18) and Eq. (21), on the conditions that $M_a = M_b$ and the stiffnesses $\chi_s, \chi_l, \chi_{ea}, \chi_{eb}$ are inversely proportional to the length of the springs,

$$\widehat{\omega}_{opt,1,2}^2(\boldsymbol{k} = \boldsymbol{0}, \chi_{ia,b} = 0) = \frac{3\sqrt{3}}{4} \widehat{\omega}_{ac,2}^2\left(k_x = \frac{2\pi}{3}, k_y = 0\right) \sim 1.3 \widehat{\omega}_{ac,2}^2(k_x = \frac{2\pi}{3}, k_y = 0). \tag{30}$$

Thus, to avoid overlapping of the neighbouring optical surfaces, the internal frequencies should be greater than the maximal acoustic frequency, $\max[\widehat{\omega}_{ac,2}^2]$ – See the Eq.'s (20) and (21)- which is caused by the additional springs in the $K_6$-system   ($\chi_{ea}, \chi_{ea} \neq 0$). One can see that the two lower optical surfaces contact/overlap each other (Fig. 11a), though at the center ($\boldsymbol{k} = \boldsymbol{0}$) these surfaces are split as opposed to results shown in Fig. 8b, 9b. The reason is the intensive repulsion between the upper acoustic surface and the lower optical surface when the ratio $m/M$ is notably greater than 1 (in the given case $\frac{m_a}{M_a} = 2.5$), and in contradistinction to the previous case the strong coupling of the external⇔the internal vibrations arise. So, in the Fig. 11a, $max(\omega_2^2(\mathbf{k})) \approx 0.44$ whereas according to the Eq. (21) $max[\widehat{\omega}_{ac,2}^2] \approx 1.15$.

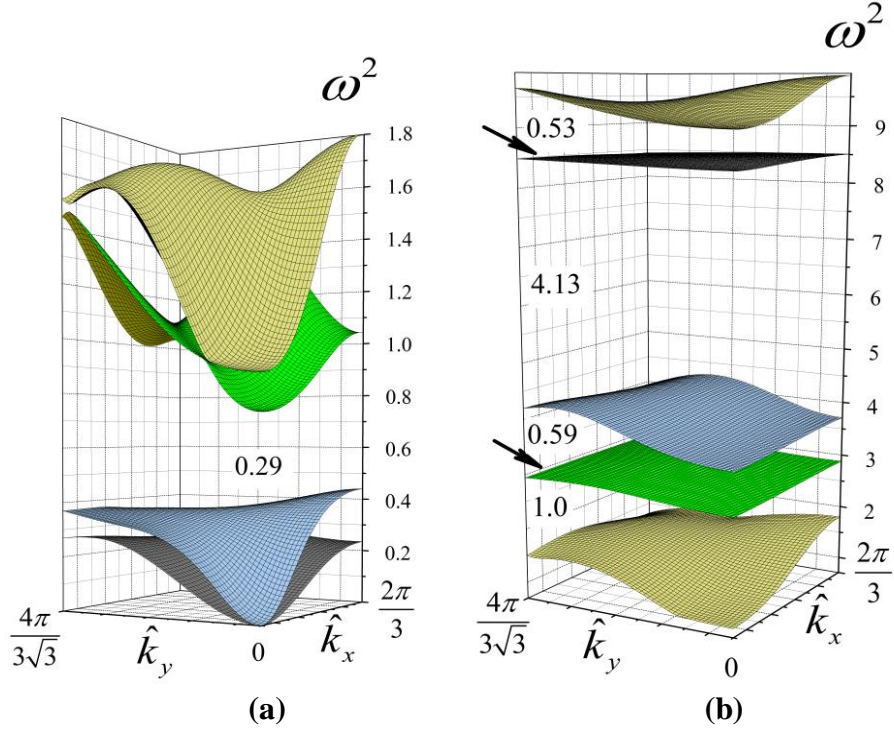

**Fig. 11**. Formation of the single optical surfaces ($K_6$-system). As above, $M_a = M_b = 2$, $\chi_s = 2/3, \chi_l = 1/3$, $\chi_{ea} = \chi_{eb} = \chi_s/\sqrt{3}$. Orientation of the internal springs within A- and B- sublattices corresponds to Fig. 3a. $m_a = 5$, $m_b = 0.1$, $\chi_{ia} = 8$, $\chi_{ib} = 8/15$; $\lambda_{xa}^2 = \lambda_{xb}^2 = 8.4$, $\lambda_{ya}^2 = \lambda_{yb}^2 = 2.8$.
a - $\omega_n^2(\mathbf{k})$, $n = 1,2,3,4$. b - $\omega_n^2(\mathbf{k})$, $n = 4,5,6,7,8$. The two arrows point the surfaces that disappear if the internal structure in B-sublattice is removed (formally, $\chi_{ib} = 0$).

Very low values of internal mass $m_b$ tempts one into thinking that in such cases masses may be neglected altogether (formally, $\chi_{ib} = 0$ in the Eq.'s (14a),(b),(c)). On the one hand, it is intuitively supposed that the morphology of some dispersion surfaces $\omega_n^2(\mathbf{k})$ practically does not alter. On the other hand, the number of independent variables in Eq. (12) decreases to six, and the dispersion equation (Eq. (13)) has only the six eigenvalues $\omega_n^2(\mathbf{k})$ instead of eight, which corresponds to the six frequency-surfaces. In reality, the disappearing surfaces are the dispersion surfaces like $\omega_5^2(\mathbf{k})$ and $\omega_7^2(\mathbf{k})$ in Fig. 11 (pointed by arrows), and in Fig. 10b where it can be seen that $\omega_5^2(\mathbf{k}), \omega_7^2(\mathbf{k})$ represent the specific surfaces that assume at $\mathbf{k} = \mathbf{0}$ the values of $\lambda_y^2, \lambda_x^2$ correspondingly (See Eq. (17)) if the condition (16) is satisfied (in Fig. 11 these surfaces are almost flat: $\omega_5^2(\mathbf{k}) \approx \lambda_y^2 = 2.8$, $\omega_7^2(\mathbf{k}) \approx \lambda_x^2 = 8.4$). So, in the discussed hypothetical case ($\chi_{ib} = 0$) the six surfaces are nearly copies of $\omega_n^2(\mathbf{k})$, $n = 1,2,3,4,6,8$ as shown in Fig. 11 and are formed through interaction between the external and internal vibrations. At the same time, the internal vibrations do not display themselves explicitly (the surfaces of the type $\omega_5^2(\mathbf{k}), \omega_7^2(\mathbf{k})$ have vanished), however; the implicit results of their existance (only in A-sublattice now) are almost the same (for the six remaining surfaces) as if both sublattices possess the internal vibration structure. Obviously, the disappearance of the two dispersion surfaces is followed by the phenomenon of broadening between remaining surfaces and widening of some band gaps.

**6.2    Dispersion surfaces of the $K_{3,3}$ system.** In this case, the additional acoustic modes of Eq. (20) are non-existent, and the condition (29) is not needed, thus the splitting between the optical surfaces can be obtained through moderate internal frequencies when, analogous to the condition (28), the average value of those is in order of $\widehat{\omega}^2_{opt,1,2}(\boldsymbol{k}=\boldsymbol{0},\chi_{ia,b}=0)$:

$$(\lambda_x^2 + \lambda_y^2)/2 \approx \widehat{\omega}^2_{opt,1,2}(\boldsymbol{k}=\boldsymbol{0},\chi_{ia,b}=0). \tag{31}$$

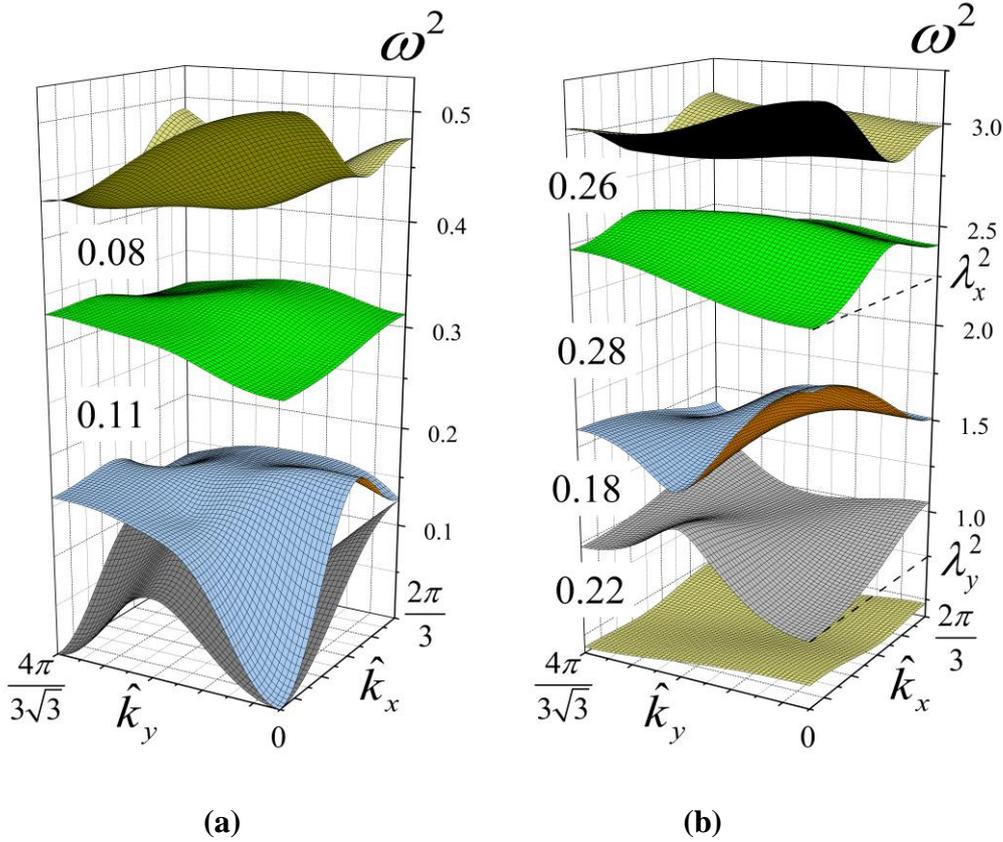

**(a)** **(b)**

**Fig. 12**.   Formation of the single optical frequency surfaces in $K_{3,3}$ system ($\chi_{ea} = \chi_{eb} = 0$). $M_a = M_b = 2$, $\chi_s = 2/3$, $\chi_l = 1/3$, $m_a = 4$, $\chi_{ia}=2$, $m_b = 1$, $\chi_{ib}=1$.
a - $\omega_n^2(\boldsymbol{k})$, $n = 1,2,3,4$; b - $\omega_n^2(\boldsymbol{k})$, $n = 4,5,6,7,8$. $\lambda_y^2 = 0.75$, $\lambda_x^2 = 2.25$, $\widehat{\omega}^2_{opt,1,2}(\boldsymbol{k}=\boldsymbol{0},\chi_{ia,b}=0) = 1.5$. The band gaps between the neighbor surfaces, $\Delta\omega^2_{n-(n+1)}$, are shown.

The corresponding result is presented in Fig. 12. There, as one can see, six the single optical surfaces emerge as the solution of the dispersion equation. The total gap size in Fig. 12, $\Delta\Omega^2$, is equal to

$$\Delta\Omega^2 = \sum \Delta\omega^2_{n-(n+1)} = 1.13, n = 2,3,....7. \tag{32}$$

It is worth to notice that for identical internal masses (either $m_a=m_b=1$, $\chi_{ia} = \chi_{ib} = 1$ or $m_a=m_b=4$, $\chi_{ia} = \chi_{ib} = 2$ – See parameters in Fig. 12) the results are much worse compared to the original result shown in Fig. 12 as far as gap formation is concerned, in spite of satisfying the ratio (31). The band gap between the upper acoustic surfaces and the lower optical surfaces disappears in both cases and for the remaining band gaps $\Delta\Omega^2 = 0.53$ and $0.74$, respectively.

As discussed above, the inequality of external masses may result in amplification of splitting between dispersion surfaces. This effect is shown in Fig. 13. Though the frequency, $\max[\omega_8^2(\mathbf{k})]$, in Fig. 13 is lower compared to that in Fig. 12, the value $\Delta\Omega^2 = 1.68$ in the first case exceeds this parameter in the second case ($\Delta\Omega^2 = 1.13$ – See Eq. (32)). The main reason for this effect is that the frequency band gap between each the optical mode and its neighbouring surface in Fig. 13 becomes about twice as narrow as in Fig. 12, as a result of heavier external masses, $M_a=5$ instead of 2.

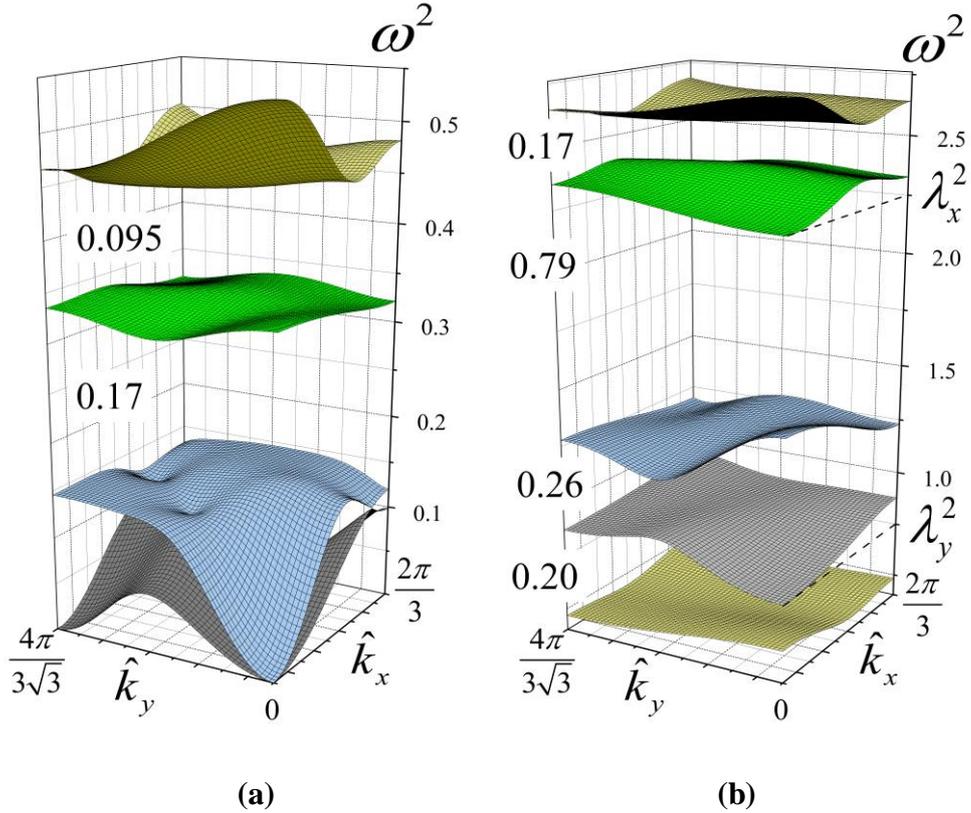

**Fig. 13.** Demonstration of the effect of inequality of the external masses. The mass $M_a=2$ (See Fig. 12) is replaced by the mass of 5, and the stiffness $\chi_{ia}=2$ by the stiffness of 10/3 so that the frequencies $\lambda_y^2$, $\lambda_x^2$ remain the same.

However, using equal external masses has its own advantages. A glimpse of the dispersion surface $\omega_6^2(\mathbf{k})$ (Fig. 12b), which is the third surface from bottom depicts the effect of this. On this surface, the derivative $\frac{d\omega}{dk}$ is negative, $\frac{d\omega}{dk} < 0$, ($k = \|\mathbf{k}\|$) over a significant portion of the Brillouin zone (See Fig. 14), which corresponds to the negative group velocity (NGV) on this part. Thus, at the corresponding $\mathbf{k}$ ($\omega_6^2(\mathbf{k}) \geq 1.6$.) the phase velocity points in the opposite direction to group velocity (direction of the wave propagation) just as it transpires in optical metamaterials. The parabolic approximations:

$$\omega_6^2(k_x, k_y = 0) = \omega_6^2(k = 0) - \left(\frac{5}{4\pi}k_x\right)^2, \tag{33}$$

$$\omega_6^2(k_x = 0, k_y) = \omega_6^2(k = 0) - \left(\frac{5\sqrt{3}}{4\pi}k_y\right)^2,$$

describe the shape of the dispersion surface and its morphology with highly accuracy within the region $\omega_6^2(\boldsymbol{k}) \geq 1.6$. The relative frequency band, $(\delta\Omega_6^2)_{NGV}$, calculated over this region is equals to:

$$(\delta\Omega_6^2)_{NGV} = \frac{2(\Omega_{6,max}^2 - \Omega_{6,min}^2)}{\Omega_{6,max}^2 + \Omega_{6,min}^2} \approx 0.21. \tag{34}$$

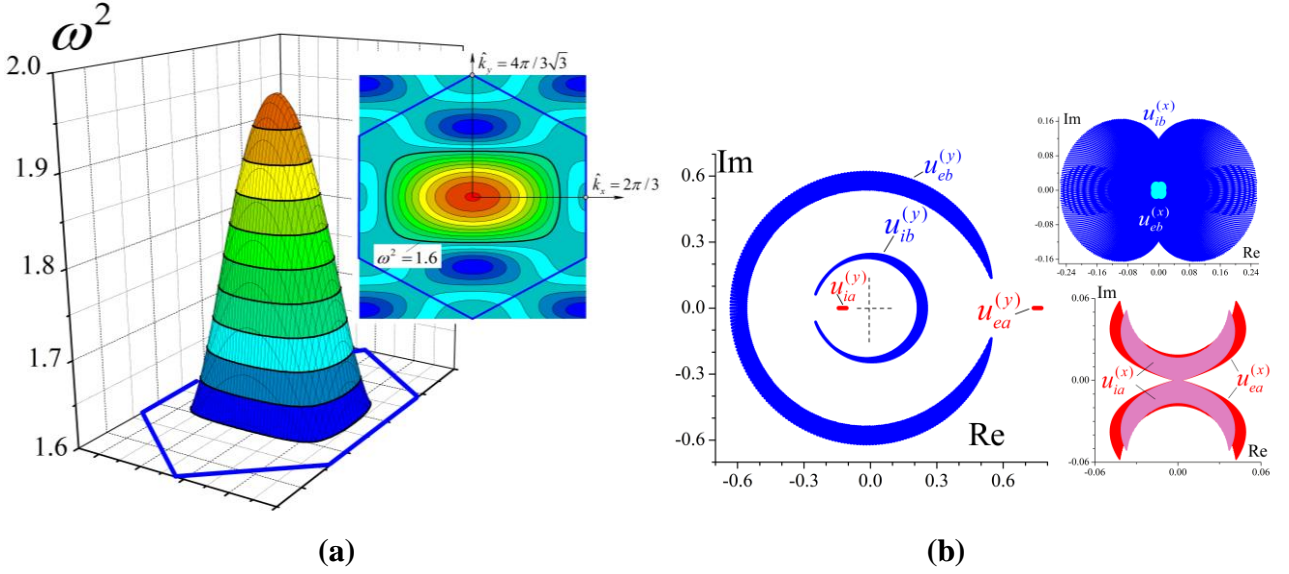

**Fig. 14.** *a* -Detailed presentation of the surface $\omega_6^2(k)$ (See Fig. 12b) at the center of the Brillouin zone, $\omega_6^2(k = 0) \approx 1.98$. *b* – The distribution of the eigen-functions, $u_{ea}^{(x)}(\boldsymbol{k}), u_{ea}^{(y)}(\boldsymbol{k}), u_{ia}^{(x)}(\boldsymbol{k})$, $u_{ia}^{(y)}(\boldsymbol{k}), u_{eb}^{(x)}(\boldsymbol{k}), u_{eb}^{(y)}(\boldsymbol{k}), u_{ib}^{(x)}(\boldsymbol{k}), u_{ib}^{(y)}(\boldsymbol{k})$, on the complex plane. The wave vector $\boldsymbol{k}$ changes within the region where $\omega_6^2(\boldsymbol{k}) \geq 1.6$. Analogously to the pair $u_{ea}^{(y)}(\boldsymbol{k}), u_{ia}^{(y)}(\boldsymbol{k})$, the complex displacements $u_{ea}^{(x)}(\boldsymbol{k})$ are in antiphase to $u_{ia}^{(x)}(\boldsymbol{k})$. This ratio is true for other pairs of the external and internal displacements.

For this type of vibration ($n = 6$, See Fig. 7a) the displacements of the external masses are mainly oriented along the y-axis, at least at $k = 0$, and $|\boldsymbol{u}_e| \gg |\boldsymbol{u}_i|$ – Fig. 7b. The result of Fig. 14b confirms this statement for the A-nodes, which include the heavy internal masses ($\frac{m_a}{M_a} = 2$). The trajectories of these nodes are largely elongated loops, and the corresponding effective mass of the A-node shells is:

$$\widehat{M}_{ay}^{(+)} \approx \frac{m_a \langle |u_{ia}^{(y)}|\rangle}{\langle u_{ea}^{(y)}\rangle} \approx m_a \times 0.17 = 0.68, \tag{35}$$

which is almost three times less than the real mass $M_a = 2$. This signifies that the energy of the A-sublattice is mainly accumulated in the phononic vibration of external masses, which may be estimated by the ratio $M_a \langle u_{ea}^{(y)}\rangle^2 / m_a \langle |u_{ia}^{(y)}|\rangle^2 \approx 17$. Unusual, at the first sight, is the result for B-nodes. The heavier external

masses vibrate with higher amplitudes than the light internal masses, $\left|u_{eb}^{(y)}\right| > \left|u_{ib}^{(y)}\right|$-Fig. 14b. Thus, the surfaces $\omega_6^2(\boldsymbol{k})$ have been formed due to availability of the internal structure at the nodes, but only a minor part of the total vibration energy is concentrated in the internal vibrations.

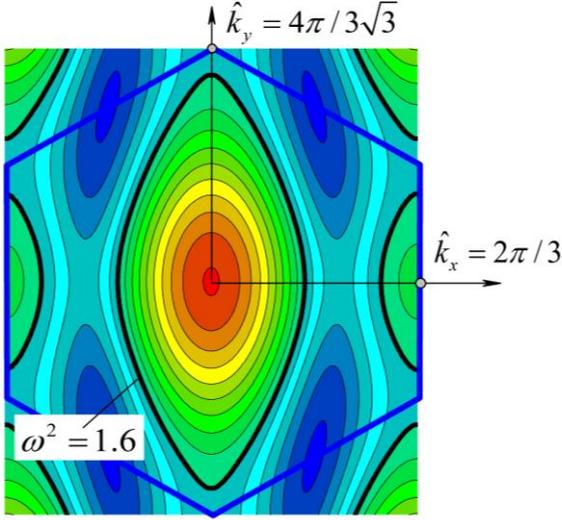

**Fig. 15.** Transformation of the surface $\omega_6^2(k)$ on turning the internal structure (in both the A- and B-sublattices) from the configuration shown in Fig. 3a to that shown in Fig. 3b. The set of other parameters is the same as in Fig. 12 and 14; $\omega_6^2(k = 0) \approx 1.98$, the gap bands, $\Delta\omega_{n-(n+1)}^2$, $n = 2,3,\ldots 7$, are equal to ~0.095, 0.095, 0.23, 0.25, 0.29 0.24 correspondingly (the total gap band $\Delta\Omega^2 \approx 1.18$).

The NGV-region at the center of the first Brillouin zone forms on the $\omega_4^2(\boldsymbol{k})$ and $\omega_8^2(\boldsymbol{k})$ dispersion surfaces as well, which can be briefly characterised as follows. On the surface $\omega_4^2(\boldsymbol{k})$ the vibration energy accumulates mainly in the internal vibrations; on the surface $\omega_8^2(\boldsymbol{k})$ – in the external vibrations just as on the surface $\omega_6^2(\boldsymbol{k})$ (but to a lesser extent than in the case of $\omega_6^2(\boldsymbol{k})$ surface). These statements may be qualitatively confirmed by the results shown in Fig. 7b on comparing the displacements of the external and internal masses at $k = 0$ for the simple case presented in that figure.

The main reason we analyze the $\omega_6^2(\boldsymbol{k})$- properties in greater detail is that this surface has both the widest frequency band, $(\delta\Omega^2)_{NGV}$-See Eq. (34), and the largest region on the BZ where $\omega^2(\boldsymbol{k})$ can be approximated by the proposed parabolic functions (See Eq.'s (33)).

Modification of the orientation of the internal structure changes the band gaps only slightly but up to some extent transforms the shape of the surface $\omega_6^2(\boldsymbol{k})$ without imposing dramatic changes to its acoustic properties (See Fig. 15 compared to Fig. 12 and Fig. 14). There are thus some properties of morphology significantly affected by the change in parameters whereas others do not alter substantially. The details of these can be derived by a close analysis of each case and scrutiny into the associated effect of alteration in each parameter as presented above.

## 7. Conclusions

The present work investigates the dependence of dispersion surface morphology and phononic band gap size in a class of metamaterials with hexagonal symmetry on topology as a primary factor. Influence of secondary parameters as heterogeneity, anisotropy and other structural attributes (hybrid vs. simple, mass ratios, etc.) is also considered. Two different topologies have been considered representing graphs of different connectivity ($K_6$ and $K_{3,3}$ systems). Due to the high number of parameters involved a limited number of configurations and parameter values have been taken into account and studied. The following conclusions have been derived:

1. To create a frequency band gap between the pairs of acoustic and optic oscillations there is no need to use different external masses for A- and B-sublattices. It would be much more effective to use the local universal mass-in-mass structure. In the case of equal external/internal masses ($M_a = M_b$, $m_a = m_b$), the isotropic internal vibrations ($\lambda_x^2 = \lambda_y^2 = \lambda^2$) form two groups of four frequency-surfaces (See Fig. 8). The lower group includes the two acoustic modes and the two lowest optic modes with negative effective masses, $\widehat{M}_{x,y}^{(-)}$. The upper group consists of the four upper optic modes, which are tangent to each other. These groups are not arbitrarily defined and are split by a considerable band gap. It is to be noted that the lower optic modes effectively suppress/decrease (especially in the $K_6$-system) the high enough acoustic modes with frequencies $\widehat{\omega}_{ac,1,2}^2$ (See Eq.'s (20)), which can be excited in A- and B-sublattices in the absence of any internal vibrations. As a result, the maximal acoustic frequency, $\max[\omega_2^2(\boldsymbol{k})]$, is much less than the possible frequency $\max[\widehat{\omega}_{ac,1,2}^2]$. Such a phenomenon can be perceived as the classical analogue of the quantum "level repulsion" and contributes to the band gap formation. For the coupling of the internal and external vibrations to be effective in this formation, the internal masses, $m_{a,b}$, must be of the order of (or higher than) the external masses, $M_{a,b}$.

2. In order to split each of the two groups of the four frequency-surfaces (discussed above) into four single pairs of acoustic/optic modes, the symmetry of the internal vibrations must be reduced by using the so called 'partial-symmetry': $\lambda_{xa}^2 + \lambda_{xb}^2 = \lambda_{ya}^2 + \lambda_{yb}^2$, $\lambda_{xa}^2 = \lambda_{yb}^2$, $\lambda_{xb}^2 = \lambda_{ya}^2$ (See Fig. 9- the A- and B-nodes are absolutely identical but possess different orientation of internal anisotropy thus different internal structure). On the one hand, detachment of the lowest optic mode from the upper acoustic mode is more expressive in the $K_6$-system (See Fig. 9b) because of additional acoustic vibrations (See Eq.'s (20)) which intensify the repulsion between the referenced modes. On the other hand, these additional vibrations broaden the upper optic surfaces, which constrict the band gaps between the neighbouring pairs.

3. Locally, there are phenomena related to the centre of the BZ and its vicinity. On further reduction of the symmetry of the internal vibrations (compared to the case of conclusion 2), when in both A- and B-nodes the internal structures are aligned along the same direction ($\lambda_{xa}^2 = \lambda_{xb}^2 \neq \lambda_{ya}^2 = \lambda_{yb}^2$), results in splitting in each of the three pairs of the optic surfaces at least at the centre of the BZ and its neighbourhood, $\boldsymbol{k} \approx 0$ (See Fig. 10). Single points of tangential contacts remaining at the periphery of the BZ can be removed by adjusting internal masses so that they are unequal: $m_a \neq m_b$ at $M_a = M_b$ and for fixed frequencies $\lambda_{xa}^2 = \lambda_{xb}^2$, $\lambda_{ya}^2 = \lambda_{yb}^2$ (See Fig.'s 11,12). In fact, this mass inequality (a specific example of heterogeneous mass distribution) magnifies the anisotropic properties of the whole system. The additional inequality of the external masses does not sufficiently change the frequency-surfaces structure.

Thus, by operating merely on the internal nodal anisotropy we can obtain different band gap structures. In particular, in the $K_{3,3}$ system the single optic surface with negative group velocity over significant part of the Brillouin zone may be realised which has several implications for possible important technical applications.